\documentclass[12pt]{article}
\pdfoutput=1
\usepackage{amsmath}
\usepackage{amssymb}
\usepackage{cite}
\usepackage{graphicx}
\usepackage{epsfig}
\usepackage{multirow}

\def\be{\begin{equation}}
\def\ee{\end{equation}}

\begin{document}
\titlepage
\begin{flushright}
IPPP/16/05  \\
\today \\
\end{flushright}

\vspace*{0.5cm}
\begin{center}
{\Large \bf  The production of a diphoton resonance via \\
\vspace*{0.5cm}
 photon--photon fusion}\\

\vspace*{1cm}
                                                   
L. A. Harland--Lang$^{1}$, V. A. Khoze$^{2,3}$, M. G. Ryskin$^{3}$ \\                                                 
                                                   
\vspace*{0.5cm}
${}^1$Department of Physics and Astronomy, University College London, WC1E 6BT, UK   \\                                                    
${}^2$Institute for Particle Physics Phenomenology, University of Durham, Durham, DH1 3LE          \\
${}^3$Petersburg Nuclear Physics Institute, NRC Kurchatov Institute, Gatchina, \linebreak[4]St. Petersburg, 188300, Russia                                              
                                                    
\vspace*{1cm} 

\begin{abstract}
\noindent Motivated by the recent LHC observation of an excess of diphoton events around an invariant mass of 750 GeV, we discuss the possibility that this is due to the decay of a new scalar or pseudoscalar resonance dominantly produced via photon--photon fusion. We present a precise calculation of the corresponding photon--photon luminosity in the inclusive and exclusive scenarios, and demonstrate that the theoretical uncertainties associated with these are small. In the inclusive channel, we show how simple cuts on the final state may help to isolate the photon--photon induced cross section from any gluon--gluon or vector boson fusion induced contribution. In the exclusive case, that is where both protons remain intact after the collision, we present a precise cross section evaluation and show how this mode is sensitive to the parity of the object, as well as potential $CP$--violating effects. We also  comment on the case of heavy--ion collisions and consider the production of new heavy colourless fermions, which may couple to such a resonance.
\end{abstract}

\end{center}

\section{Introduction}
The ATLAS and CMS collaborations have recently reported the observation of an intriguing excess of events in the diphoton mass distribution around 750 GeV~\cite{ATLAS:2015exc,CMS:2015dxe} in the $\sqrt{s}=13$ TeV Run--II data. This has stimulated a great deal of theoretical interest, and the possibility of interpreting this observation in terms of a new 750 GeV resonance, $R$, decaying to two photons has been considered within a wide range of models. One interesting, and in some sense natural possibility, given that this possible excess is only seen so far in the $\gamma\gamma$ decay channel (and not, for example, in the dijet mass spectrum), is that the resonance may couple dominantly to photons, with the coupling to gluons and other coloured particles being either suppressed or absent entirely. This has been discussed in~\cite{Fichet:2015vvy,Csaki:2015vek} (see also~\cite{Anchordoqui:2015jxc,Franceschini:2015kwy,Danielsson:2016nyy,Nomura:2016fzs,Ito:2016zkz,D'Eramo:2016mgv,Csaki:2016raa,Fichet:2016pvq,Abel:2016pyc,Barrie:2016ntq,Ben-Dayan:2016gxw}), where the $\gamma\gamma$ induced process in both the usual inclusive and the exclusive modes has been considered. In the latter case the incoming photons are emitted coherently from the colliding protons, which then remain intact after the collision; this will naturally occur in a non--negligible fraction of events.

In light of this, we present in this paper a precise theoretical determination of both the inclusive and exclusive $\gamma\gamma$ luminosities which are required to calculate the corresponding production cross sections. We will assume this excess is due to the decay of a scalar or pseudoscalar resonance $R$ of mass $M_R=750$ GeV, although of course it remains possible that it is due to a simple statistical fluctuation. In general it is not the aim of this paper to make further model--dependent assumptions about the nature of the resonance, but rather to give the most accurate possible calculation of the relevant $\gamma\gamma$ luminosities for a range of kinematic conditions, improving upon the more phenomenological approaches taken in e.g.~\cite{Fichet:2015vvy,Csaki:2015vek}. While these give a reasonable estimate of the expected rates, at the $\sim 10-50\,\%$ level, a more precise calculation may be needed for future higher statistics studies, provided the excess survives. 

Thus, in this paper we account for the NLO in $\alpha_s$ corrections to the photon PDF evolution, and demonstrate that the inclusive luminosity relevant to the production of such a high mass object is under good theoretical control, up to a $\sim \pm 15-20\,\%$ uncertainty. This then allows the branching ratio for $R\to \gamma\gamma$ to be determined quite precisely, if the $\gamma\gamma$--induced production mechanism is indeed dominant, provided the cross section and total width are measured. Under this assumption, the lack of observed excess at $\sqrt{s}=8$ TeV  allows a quite precise limit to be placed on the corresponding cross section at $\sqrt{s}=13$ TeV. We also consider the possibility that the resonance may be produced in gluon--gluon and vector boson fusion (VBF), and show quantitatively how relatively simple final--state cuts, namely vetoing on additional jets close in rapidity to the resonance and selecting events where the resonance has a reasonably low transverse momentum, respectively, are expected to suppress these contributions relative to the photon--induced case.

In addition, we consider the possibility for producing this resonance in central exclusive production (CEP), with two intact protons in the final--state. The experimental situation at the LHC is very encouraging here, and such processes may be measured with both protons tagged using the approved and installed AFP~\cite{CERN-LHCC-2011-012,Tasevsky:2015xya} and CT--PPS~\cite{Albrow:1753795} forward proton spectrometers, associated with the ATLAS and CMS central detectors, respectively, see also~\cite{yp}. Moreover, the measurement of new heavy objects with tagged protons can in general be highly advantageous, see~\cite{Khoze:2001xm,FP420,Heinemeyer:2007tu,deFavereaudeJeneret:2009db,Chapon:2009hh,Heinemeyer:2010gs,Fichet:2014uka}. We therefore present a precise evaluation of the exclusive $\gamma\gamma$ luminosity, accounting for all physical effects, including the probability of no additional underlying event activity, or so--called survival factor. For photon--mediated processes, this turns out to be quite close to unity, but it cannot be ignored; the treatment of this latter factor is either absent or approximate in other studies such as~\cite{Fichet:2015vvy,Csaki:2015vek}. The uncertainty is found to be very small, at the percent level, allowing the corresponding exclusive cross section to be accurately predicted.

For the exclusive production of such a high mass object there is naturally a strong suppression in the gluon--initiated channel, where the perturbative probability for producing no extra particles is strongly damped, compared to the $\gamma\gamma$ one (see~\cite{Khoze:2001xm} for the first discussion of this),  while the heavy bosons exchanged in VBF inevitably lead to proton dissociation. This exclusive mode therefore naturally leads to a relative enhancement in the $\gamma\gamma$--induced process. In addition, it offers the advantage that the resonance quantum numbers, as well as potential CP--violating effects~\cite{Khoze:2004rc}, may be determined, through measurements of the azimuthal correlations between the transverse momenta of detected protons.  We discuss these possibilities in detail here, and show how a scalar or pseudoscalar state leads to dramatically different proton distributions, so that with only a handful of events these possibilities may be distinguished. We also comment on the potential for producing such a resonance exclusively in heavy ion collisions, and provide estimates for the inclusive  $\gamma\gamma$--induced cross section for the production of new heavy colourless fermions which could be associated with the resonance.

The outline of this paper is as follows. In Section~\ref{sec:gaminc} we present the theoretical ingredients required to calculate the inclusive $\gamma\gamma$ luminosity and discuss the implications for the 750 GeV excess. In Section~\ref{sec:pdf} we compare our results to the predictions from other photon PDF sets. In Section~\ref{sec:vbf} we consider the possibility of $gg$ and $WW$ initiated contributions to the production of the resonance $R$ and calculate the impact of relatively simple cuts on the final--state, showing how these can suppress these components relative to  the $\gamma\gamma$--initiated case. In Section~\ref{sec:events} we comment on the event structure in $\gamma\gamma$--initiated processes. In Section~\ref{sec:CEP} we consider the purely exclusive $\gamma\gamma$--initiated production mechanism, where the protons remains intact after the collision, and show how the corresponding $\gamma\gamma$ luminosity may be calculated. In Section~\ref{sec:spin} we calculate the distributions with respect to the transverse momenta of the outgoing protons, and demonstrate how these are strongly sensitive to the parity of the produced resonance. In Section~\ref{sec:cp} we discuss how a measurement of any asymmetry in such a distribution may in addition be sensitive to CP--violating effects in the production mechanism. In Section~\ref{sec:ion} we comment briefly on the possibility to produce such a resonance exclusively in heavy ion collisions, and show that this is not likely to be a viable channel. In Section~\ref{sec:fermion} we consider the case of inclusive heavy colourless fermion production. Finally, in Section~\ref{sec:conc} we conclude.

\section{Inclusive production}

\subsection{$\gamma\gamma$ initiated cross section}\label{sec:gaminc}

To calculate the cross section, $\sigma^{\rm inc}(pp \to R)$, for the inclusive production of a scalar (or pseudoscalar) resonance, we first define the usual inclusive $\gamma\gamma$ luminosity for the production of a system $X$ of mass $M_X$ and rapidity $y_X$, which is given by
\be\label{lumiinc}
\frac{{\rm d}\mathcal{L^{\rm inc}_{\gamma\gamma}}}{{\rm d}M_X^2\,{\rm d}y_X}=\frac{1}{s}\gamma(x_1,\mu)\,\gamma(x_2,\mu)\;,
\ee
where  $x_{1,2}=\frac{M_X}{\sqrt{s}}e^{\pm y_X}$ are the proton momentum fractions carried by the photons and $\gamma(x,\mu)$ is the photon parton distribution function (PDF), i.e. the density of the photons with  momentum fraction $x$ at the scale $\mu\sim M_X$. In terms of this, the inclusive cross section for $\gamma\gamma \to X$ is 
\be
\frac{{\rm d}\sigma^{\rm inc}(pp\to X)}{{\rm d}M_X^2\, {\rm d}y_X}=\frac{{\rm d}\mathcal{L^{\rm inc}}}{{\rm d}M_X^2\,{\rm d}y_X} \,\hat{\sigma}(\gamma\gamma\to X)\;,
\ee
were $\hat{\sigma}$ is the cross section for the $\gamma\gamma \to X$ subprocess. If we consider the production of a resonance $R$ of mass $M_R$ and rapidity $y_R$ then in the narrow width approximation\footnote{If the resonance width is large enough, then this approximation is not completely valid. As the $\gamma\gamma$ luminosity is steeply falling with $M_X$, the distribution will be fairly slowly convergent for $M_X<M_R$, so that for e.g. $\Gamma_{\rm tot}=45$ GeV, the predicted cross section is $\sim 10\%$ larger than in the narrow width case. If such a large width persists in future higher precision data, than this should be taken into account consistently.}  the subprocess cross section is 
\begin{align}
\hat\sigma(\gamma\gamma\to R)&=\frac{8\pi^2\Gamma(R\to\gamma\gamma)}{M_R}\delta(M_R^2-M_X^2)\;,\\
&=
\frac{8\pi^2\Gamma_{\rm tot}(R)}{M_R}\,{\rm Br}(R\to\gamma\gamma)\,\delta(M_R^2-M_X^2)\; ,
\label{h-incl}
\end{align}
and thus 
\be
\frac{{\rm d}\sigma^{\rm inc}(pp\to R)}{{\rm d}y_R}= \frac{8\pi^2\Gamma(R\to\gamma\gamma)}{M_R}\,\frac{{\rm d}\mathcal{L}^{\rm inc}_{\gamma\gamma}}{{\rm d}y_R\,{\rm d}M_X^2}\bigg|_{M_X=M_R} \;.
\label{s-incl}
\ee
The photon PDF is given in terms of an input term $\gamma(x,Q^2_0)$ at the starting scale $Q_0$, and a term due to  photon emission from quarks during the DGLAP evolution from $Q^2_0$ to $Q^2$.  The input  $\gamma(x,Q^2_0)$  may be written in terms of a coherent component, $\gamma^{\rm coh.}(x,Q_0)$, due to the elastic process, $p\to p+\gamma$, see~\cite{Martin:2014nqa}, as well as an incoherent component,  $\gamma^{\rm incoh.}(x,Q_0)$, due to emission from the individual quarks within the proton (i.e. the direct analogue of perturbative emission in the QCD case). In the former case we take the more precise form given by (\ref{WWflux}) for the coherent photon flux, rather than the approximate expression used in~\cite{Martin:2014nqa}. The coherent component gives the dominant contribution at the input scale $Q_0$, with $\gamma^{\rm coh.}(x,Q_0)/\gamma^{\rm incoh.}(x,Q_0)\approx 3$, with some $\sim 10\%$ variation in this depending on the precise value of $x$.

For the DGLAP evolution, since the QED coupling $\alpha$ is very small it is sufficient to consider just the leading $O(\alpha)$ contribution to the photon PDF, $\gamma(x,Q^2)$, although we account for the running of $\alpha$ to 1--loop order, as the relevant scale $\mu \sim M_R$ is quite large. The appropriate splitting functions which allow the evolution to be evaluated at NLO in the strong coupling $\alpha_S$ have recently been calculated in~\cite{deFlorian:2015ujt}, and are included here\footnote{Strictly speaking, to be consistent we should also include the $\gamma\gamma\to R$ matrix element at NLO, however if the experimental value of the $R\to \gamma\gamma$ width is taken this implicitly includes higher order--QCD corrections, while for the simplest case that $R$ does not couple to coloured particles these corrections are zero.}.
Thus, we have
\begin{align}\nonumber
\gamma(x,\mu^2)&=\gamma(x,Q_0^2)+\int_{Q_0^2}^{\mu^2}\frac{\alpha(Q^2)}{2\pi}\frac{dQ^2}
{Q^2}\int_x^1\frac{dz}z \bigg(P_{\gamma\gamma}(z)\gamma(\frac xz,Q^2)\\ \label{pdf}
&+\sum_q e^2_qP_{\gamma q}(z)q(\frac xz,Q^2)+ P_{\gamma g}(z)g(\frac xz,Q^2)\bigg)\;,
\end{align}
where the input distribution $\gamma(x,Q_0)=\gamma^{\rm coh}(x,Q_0)+\gamma^{\rm incoh}(x,Q_0)$ and $P_{\gamma q}(z)$ and $P_{\gamma g}(z)$ are the NLO (in $\alpha_S$) splitting functions. At LO we have
 \begin{align}
 P_{\gamma g}(z)&=0\; , \\
 P_{\gamma q}(z)&=\left[\frac{1+(1-z)^2}z\right]\;,\\
 P_{\gamma\gamma}(z)&=-\frac{2}{3}\left[N_c\sum_q e^2_q +\sum_l e^2_l\right]\delta(1-z)\;,
 \end{align}
 where the indices $q$ and $l$ denote the light quark and the lepton flavours respectively, see~\cite{deFlorian:2015ujt} for the full NLO results. We find that including the NLO form of the DGLAP evolution reduces the predicted cross section for $M_R=$ 750 GeV by about $5\%$ compared to LO, with the suppression being slightly larger at the highest rapidities. 

What are the uncertainties on the above expressions? The main source is in fact due to varying the factorization scale in the photon PDF, indicating the potential importance of higher--order contributions. Varying $\mu_R$ (in $\alpha$ and $\alpha_s$) and $\mu_F$ independently between $(M_R/2,2M_R)$ for $M_R=750$ GeV, we find that there is a $\sim \pm 10\,\%$ variation in the predicted $\gamma\gamma$ luminosity, and hence in the predicted inclusive cross section. This is dominantly due to the factorization scale variation, while if we set $\mu_R=\mu_F$ some compensation in fact occurs, so that the variation is instead $\sim 5$\%.  There is also some error associated with the PDF uncertainty of the quark and gluon PDFs which enter the photon DGLAP evolution. Here, we take MMHTNLO~\cite{Harland-Lang:2014zoa} PDFs\footnote{Strictly speaking, a set which includes the photon PDF in the fit should be used, however an up--to--date fit within the framework described in this paper is not currently available, and moreover this will only influence the PDFs at higher order in $\alpha$, so will be a small effect.}: calculating the PDF uncertainty in the usual way we find less than a $\sim \pm 2\%$ variation.

In addition there is some uncertainty due to the quark treatment in the `incoherent' emission term in the input PDF $\gamma(x,Q^2_0)$, and the related question of the choice of starting scale $Q_0$, which acts as an upper limit on the scale for photon emission in both the coherent and incoherent input components; here we take $Q_0=1$ GeV. We choose to freeze the quark PDFs below the starting scale $Q<Q_0$ at $Q_0$, corresponding to an upper limit on the incoherent term (see~\cite{Martin:2014nqa} for more details). The remaining contribution from the coherent component for $Q>Q_0$ is included by adding a corresponding term to the photon PDF evolution. However, this is not the only way that the incoherent input component may be treated, so to give a rough estimate of the uncertainty associated with this we can simply set $\gamma^{\rm incoh}(x,Q_0)=0$: in this case the resonance $R$ production cross section decreases by $\sim 15$\%. Clearly this represents an extreme and physically unjustified choice, so more realistically we can expect the uncertainty to be smaller than this. Thus combining this with the scale variation and other sources, we expect the total uncertainty to be of order $\sim \pm 15-20$\%. We note however that further studies to constrain the incoherent component of the photon PDF, in the context of a global PDF fit, can reduce the uncertainty associated with this.

\begin{figure}
\begin{center}
\includegraphics[scale=0.65]{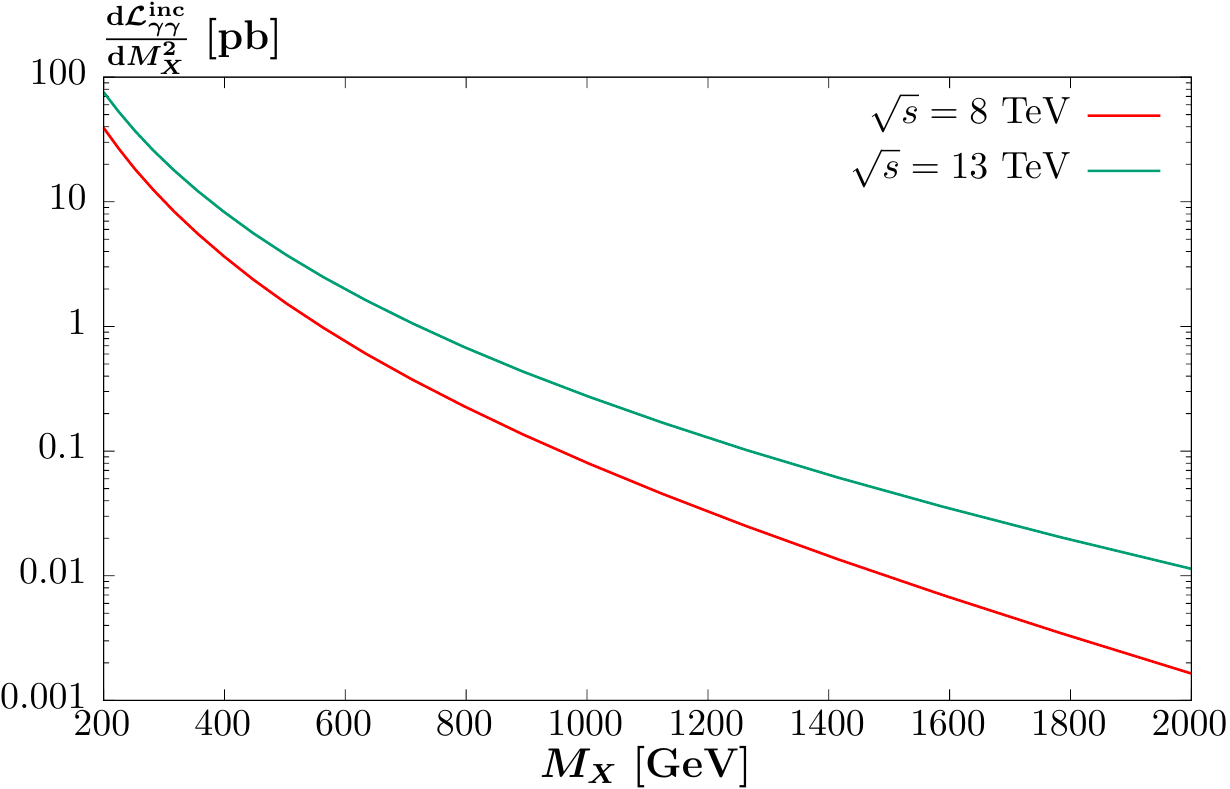}
\includegraphics[scale=0.65]{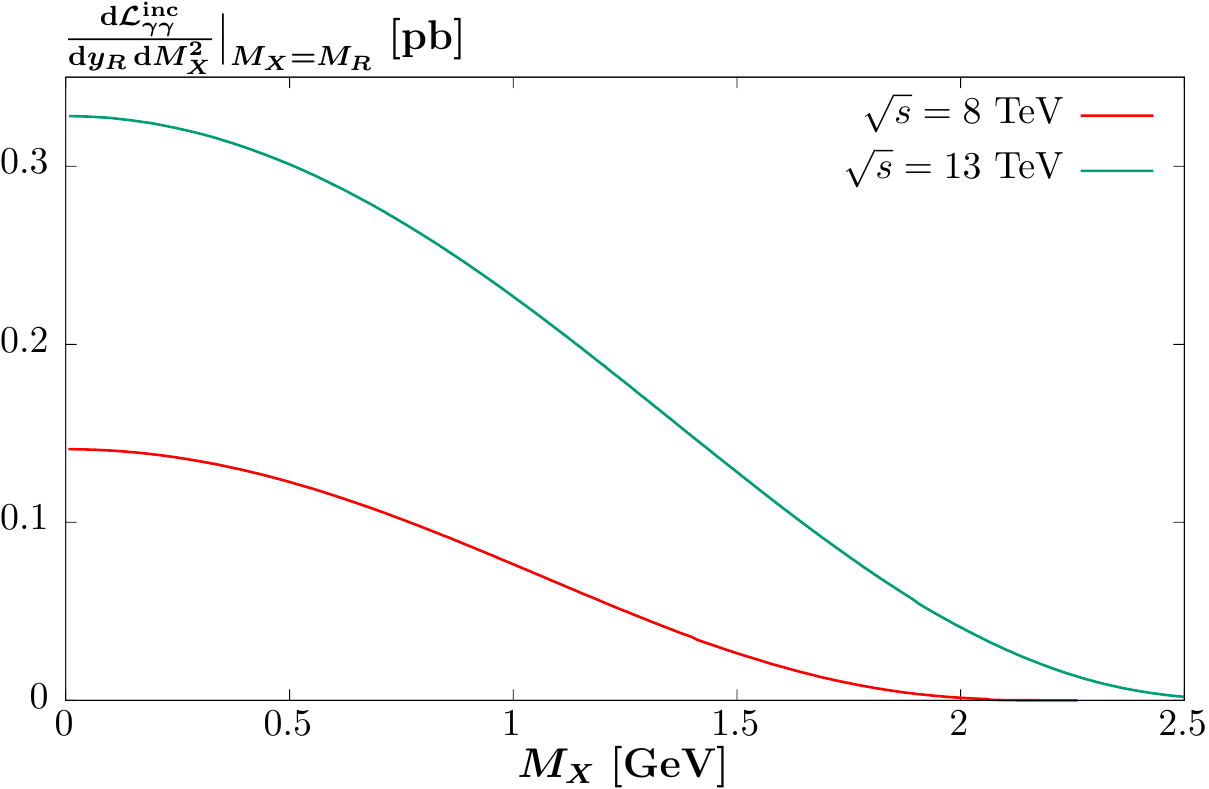}
\caption{Inclusive $\gamma\gamma$ luminosity at $\sqrt{s}=8$ and 13 TeV, shown (left) differential in the invariant mass, $M_X$, of the produced system, integrated over rapidity $y_X$, (right) for the production of a resonance of mass $M_R=750$ GeV, differential in the particle rapidity.}\label{fig:lumi}
\end{center}
\end{figure}

The inclusive $\gamma\gamma$ luminosity as a function of the system invariant mass $M_X$ is shown in Fig.~\ref{fig:lumi} (left), at both $\sqrt{s}=8$ and 13 TeV. The luminosity ratio of the higher to the lower energy increases significantly at higher $M_X$, due to the higher parton $x$ values probed, from $\sim 2-10$ over the mass range (200--2000 GeV) considered. At $M_X=750$ GeV we find
\be\label{lumirat}
\frac{\mathcal{L}^{{\rm inc}}_{\gamma\gamma}(\sqrt{s}=13\,{\rm TeV})}{\mathcal{L}^{{\rm inc}}_{\gamma\gamma}(\sqrt{s}=8\,{\rm TeV})}=2.9\;,
\ee
as can be seen in Fig.~\ref{fig:lumi} (right), where the luminosities for the production of a $M_R=750$ GeV resonance as a function of the resonance rapidity $y_R$ is also shown. Thus, if for illustration we take the $\sqrt{s}=13$ TeV production cross section of $\sigma^{\rm  inc}=4-8$ fb, as indicated in e.g.~\cite{Ellis:2015oso,Djouadi:2016eyy} we have
\be
\sigma^{\rm inc}_{8\,{\rm TeV}}(pp \to (R\to\gamma\gamma))=1.4-2.8\,{\rm fb}\;.
\ee
The lack of excess seen at $\sqrt{s}=8$ TeV in the $\gamma\gamma$ mass spectrum roughly implies
that $\sigma<2-3$ fb~\cite{Khachatryan:2015qba}. Phrased differently, if we take this as our limit, then if the measured cross section at $\sqrt{s}=13$ TeV exceeds about $6-9$ fb, the hypothesis that the resonance is dominantly produced in $\gamma\gamma$ fusion becomes ruled out.

Considering the cross section for resonance production, from (\ref{h-incl}) and our results for the $\gamma\gamma$ luminosity, we get the simple relation (for $M_R=750$ GeV)
\be
\sigma^{\rm inc}(pp\to (R\to \gamma\gamma))=91\,{\rm fb}\, \left(\frac{\Gamma_{\rm tot}(R)}{1\,{\rm GeV}}\right){\rm Br}(R\to \gamma\gamma)^2\;,
\ee
or, rearranging
\be
{\rm Br}(R\to \gamma\gamma)=\frac{1}{9.5}\left(\frac{\sigma^{\rm inc}[{\rm fb}]}{\Gamma_{\rm tot}(R)/1\,{\rm GeV}}\right)^{1/2}\;,
\ee
where in the latter case the inclusive $R\to \gamma\gamma$ cross section is given in fb. Thus, if as above for illustration we take the production cross section of $\sigma^{\rm  inc}=4-8$ fb as indicated by e.g.~\cite{Djouadi:2016eyy}, as well as the width $\Gamma_{\rm tot}=45$ GeV, then we find
\be 
{\rm Br}(R\to \gamma\gamma)=3.1-4.4\, \%\;.
\ee
Clearly this low branching ratio, if correct, indicates that such a new resonance must have a sizeable branching into other SM (most notably, $W/Z$ modes, which should be present) or BSM particles. Of course, the evidence for such a high total width, which is preferred by the ATLAS but not the CMS data, is at this stage only tentative, and the true width may be lower. As an extreme, if we assume that the resonance only couples to photons, i.e. ${\rm Br}(R\to \gamma\gamma)=100\,\%$, then we expect 
\be
\Gamma_{\rm tot}=44-88\, {\rm MeV}\;,
\ee
for the cross section range  $\sigma^{\rm  inc}=4-8$ fb.

\subsection{Comparison to other PDF sets}\label{sec:pdf}

We have so far presented the results for resonance production within the approach set out in~\cite{Martin:2014nqa} (see also~\cite{Harland-Lang:2016apc}). However, other predictions for the photon PDF are available, in particular in the context of the major global parton analyses, namely the NNPDF2.3QED~\cite{Ball:2013hta}, CT14QED~\cite{Schmidt:2015zda} and the older MRST2004QED~\cite{Martin:2004dh} sets. In the NNPDF2.3 QED parton analysis, the photon PDF is freely parameterised in the same way as for other partons. This is then fitted to DIS and a small set of LHC data, namely $W,Z$ and high/low--mass Drell--Yan production (more precisely this is achieved by Bayesian reweighting, see~\cite{Ball:2013hta} for full details). Unfortunately such data currently has limited constraining power, and the resulting PDF uncertainties are very large, in particular at higher $x$; we will see this below in Table~\ref{table:lumi}. The MRST2004 and CT14 QED sets take a different approach, which is more similar to that considered here: they both assume a theoretical form due to photon emission from the individual (valence) quarks within the proton, i.e. equivalent to the incoherent input component in this work. In the MRST2004 case two sets are available, corresponding to whether current or constituent quark masses are used in the input. In the CT14 set an additional free parameter, given in terms of the momentum fraction, $p^\gamma_0$, carried by the input photon PDF, is introduced, and by comparing to the ZEUS $e p\to e\gamma X$ data~\cite{Chekanov:2009dq}, is found to be 0 -- 0.14\% for the starting scale $Q_0= 1.295$ GeV at 90\% confidence.
 
\begin{table}
\begin{center}
\def\arraystretch{1.4}
\begin{tabular}{|c|c|c|c|c|}
\hline
&CT14 ($p_0^\gamma=0-0.14$\%) &MRST2004&NNPDF2.3&This work \\ \hline
$\mathcal{L}^{{\rm inc}}_{\gamma\gamma}(13\, {\rm TeV})/\mathcal{L}^{{\rm inc} }_{\gamma\gamma}( 8 \,{\rm TeV})$&3.1 -- 2.8 &$2.65\pm 0.15$&$2.1\pm 0.4$&2.9\\
\hline
$\mathcal{L}^{{\rm inc}}_{\gamma\gamma}( {\rm PDF})/\mathcal{L}^{{\rm inc}}_{\gamma\gamma}( {\rm Sect.}~\ref{sec:gaminc})$&0.4 -- 0.9 &$1.1\pm0.5$&$2.1\pm1.4$\\
\cline{1-4}
\end{tabular}
\caption{Ratio of the inclusive $\gamma\gamma$ luminosities at $M_X=750$ GeV, defined in (\ref{lumiinc}), at $\sqrt{s}=13$ TeV to 8 TeV, for a range of QED  PDF sets, described in the text, compared to the central prediction (\ref{lumirat}) from this work, and the ratio of the $\gamma\gamma$ luminosities at $\sqrt{s}=13$ TeV for these sets to the prediction of this paper. The MRST2004 range corresponds to the constituent and current quark mass results, the CT14 range to the results with the photon momentum fraction $p_0^\gamma$ between 0 -- 0.14\%, and the NNPDF2.3 uncertainties correspond to a 68\% confidence envelope.}
\label{table:lumi}
\end{center}
\end{table}

In Table~\ref{table:lumi} we compare our prediction (\ref{lumirat}) for the ratio of the inclusive $\gamma\gamma$ luminosities for a 750 GeV resonance at $\sqrt{s}=13$ TeV to 8 TeV to the results of these PDF sets, including their corresponding uncertainties, calculated as described in the table caption.  In light of the lack of an observation of any excess in the diphoton channel at 8 TeV, this is an important quantity: as discussed in the preceding section, we expect the ratio to be $\sim 3$, which is sufficiently large that the current best fit cross sections corresponding to the excess at 13 TeV are not in strong tension with this result. We find that generally the predicted ratios in Table~\ref{table:lumi} are consistent with our results, with the exception of the NNPDF2.3 set, which predicts a somewhat lower ratio. Thus those sets which include theoretical guidance for the photon PDF appear to prefer a somewhat higher value of this ratio, compared to the NNPDF case (this is also clear from Fig. 11 of~\cite{Schmidt:2015zda}, where the NNPDF set shows a flatter behaviour with decreasing $x$). However, when uncertainties are accounted for the tension is not too dramatic. In other works (see e.g.~\cite{Franceschini:2015kwy,Csaki:2016raa}) a somewhat lower value of $\sim 2$ is quoted, however it is important in this case to account for the uncertainties in such a prediction. Moreover, it should be pointed out that such a low central value is only found in the case of the NNPDF set, and is not preferred by analyses which use theoretical guidance or, in the case of this work, all available experimental input in the form of the contribution from coherent photon emission, to constrain the photon PDF.
  
 Also shown in Table~\ref{table:lumi} is the ratio of the $\gamma\gamma$ luminosity for a 750 resonance at $\sqrt{s}=13$ predicted by these PDF sets  to our central prediction. Here, the results are completely consistent within uncertainties with ours, with the slight exception of the CT14 set, which appears to prefers a slightly lower value (recalling that the quoted intervals correspond to 90\% confidence limits). However, in the CT14 set the photon momentum fraction $p^\gamma_0$ is largely insensitive to the precise form of the photon PDF in the $x$ region relevant to 750 GeV resonance production ($x\sim 0.09$ (0.06) at $\sqrt{s}=8$ (13) TeV), where it is small in size. In~\cite{Schmidt:2015zda}, the form of the photon PDF in this  $x$ region is driven by the assumption of a purely incoherent input, while the ZEUS data~\cite{Chekanov:2009dq} which they fit to does not directly constrain the photon PDF for $x$ above $\sim 0.02$. Thus, while for our set we find $p_0^\gamma=0.2\,\%$ at the CT14 starting scale, any apparent tension here and in the predicted luminosity should not be taken literally. The very large uncertainty in the NNPDF prediction  is also evident.

Although we have for the purposes of presenting a full discussion compared our results with the best available alternative photon PDF sets, we would argue that these approaches miss a crucial element of the physics involved with the photon PDF. In particular, we have seen in Section~\ref{sec:gaminc} that at the starting scale $Q_0$ the dominant contribution to the photon PDF is generated by coherent emission of a photon due to the electric charge of the entire proton, a theoretically well understood and experimentally well constrained process. The reason this applies here, and not in the case of the PDFs of the quarks and gluons, is that QED corresponds to a long range force that does not suffer from the issue of non--perturbativity at low scales. Thus a significant part of the experimental input, which must be provided by a global analysis for the case of the quark/gluon PDFs, is already present for the photon PDF, through the measurement of the proton form factors (see (\ref{WWflux}) below) for coherent photon emission; in QCD there is no equivalent to this. Indeed, the coherent emission process has been observed experimentally at the LHC, e.g. in the ATLAS measurement~\cite{Aad:2015bwa} of exclusive lepton pair production at $\sqrt{s}=7$ TeV, with results that are in good agreement with the expectations for coherent photon--initiated production, once all relevant effects (i.e. both the survival factor and the Sudakov factor) are accounted for; such a component must certainly be present in inclusive processes as well. The remaining fairly small component due to incoherent photon emission from the individual quarks may be effectively modelled down to very low scales, and a $\sim 10-15\%$ uncertainty may be assigned due to this, even without further experimental input which can be provided by including the photon PDF in a global parton analysis. These arguments are particularly relevant for the production of a 750 GeV resonance, where for the $x$ values relevant to this process the uncertainties associated with any set which omits this coherent component are currently large.

\subsection{VBF and coloured particle fusion processes}\label{sec:vbf}

As well as coupling to photons, we will in general expect the resonance $R$ to couple to $W$ and $Z$ bosons, and so to be produced by vector boson fusion (VBF). In addition, if it couples to colour, we may expect the $gg\to X$ process to contribute to the production cross section. Moreover, even if this is not the case, the irreducible background from continuum diphoton production proceeds primarily through the $q\overline{q}$ channel; although we do not consider this explicitly here, the results for the $gg$ channel will be qualitatively similar, as initial--state gluons act as a significant source of quarks in their DGLAP evolution. In order to separate these from the $\gamma\gamma$ induced signal process, additional cuts can be placed. In particular, to suppress the VBF contribution a fairly low cut can be imposed on the resonance transverse momentum $p_\perp^R<p_\perp^c\ll M_W$, which will reduce the VBF cross section by a factor $\sim (p_\perp^c)^2/M^2_W\ll 1$. To suppress the $gg$ initiated contribution, we can veto on events with jets of transverse momentum greater than some cut, $k_\perp>k_\perp^c$ in a rapidity interval $\delta \eta$ on both sides of the resonance; in this case the $gg$ fusion process will generally be accompanied by the  bremsstrahlung of additional high $k_\perp$ gluons (see e.g.~\cite{Csaki:2016raa} for a MC study). In both cases, the cuts $p_\perp^c$ and $k_\perp^c$ should be chosen to be sufficiently large that the underlying event does not generally produce activity passing these cuts.

To calculate the effect of these cuts (which are analogous to those used to select Higgs boson production via VBF~\cite{Chehime:1992ub,Rainwater:1997dg}) on the $\gamma\gamma$ cross section, we can apply the simple approach described in~\cite{Harland-Lang:2016apc}. Namely, we should limit the phase space region for the splitting functions $P_{\gamma q}$ and $P_{\gamma g}$, corresponding to real emission in the DGLAP evolution (\ref{pdf}) of the photon PDF, to remove the case that the final--state partons are radiated into the `veto' interval. Due to the strong $k_\perp$ and angular ordering of the DGLAP evolution it is sufficient to include this constraint in the last step of evolution only; if in this step the vetoes are satisfied then all partons emitted in previous steps in the DGLAP ladder will automatically satisfy them. From a simple consideration of the kinematics of the final splitting, we find that these vetoes are imposed by adjusting the $P_{\gamma q}(z)$ and $P_{\gamma g}(z)$ functions so that if
\be\label{gg-supp}
1-z<\frac{k_\perp}{k_\perp+M_R e^{-\delta\eta}}\;,\qquad k_\perp>k_\perp^c\;,
\ee
in  the $gg$ induced process and
\begin{equation}
p_\perp^R>p_\perp^c\;,
\end{equation}
in the VBF case, they are set to zero. More precisely, in the case of the NLO splitting functions we should consider vetoes on the two emitted partons individually, i.e. $qg$($q\overline{q}$) for  $P_{\gamma q}$($P_{\gamma g}$). However since the effect of the NLO correction is rather small ($\sim 5$\% ) here we for simplicity use the same veto as in the LO case. This corresponds to a veto on the kinematics of the parton pair and so only gives an approximate indication of the effect to the NLO contribution; for example the transverse momentum cut will be overly restrictive. This introduces an additional sub--percent level uncertainty in the calculation, which is however well within other uncertainties due to the scale choice and input photon PDF.

\begin{figure}
\begin{center}
\includegraphics[scale=0.65]{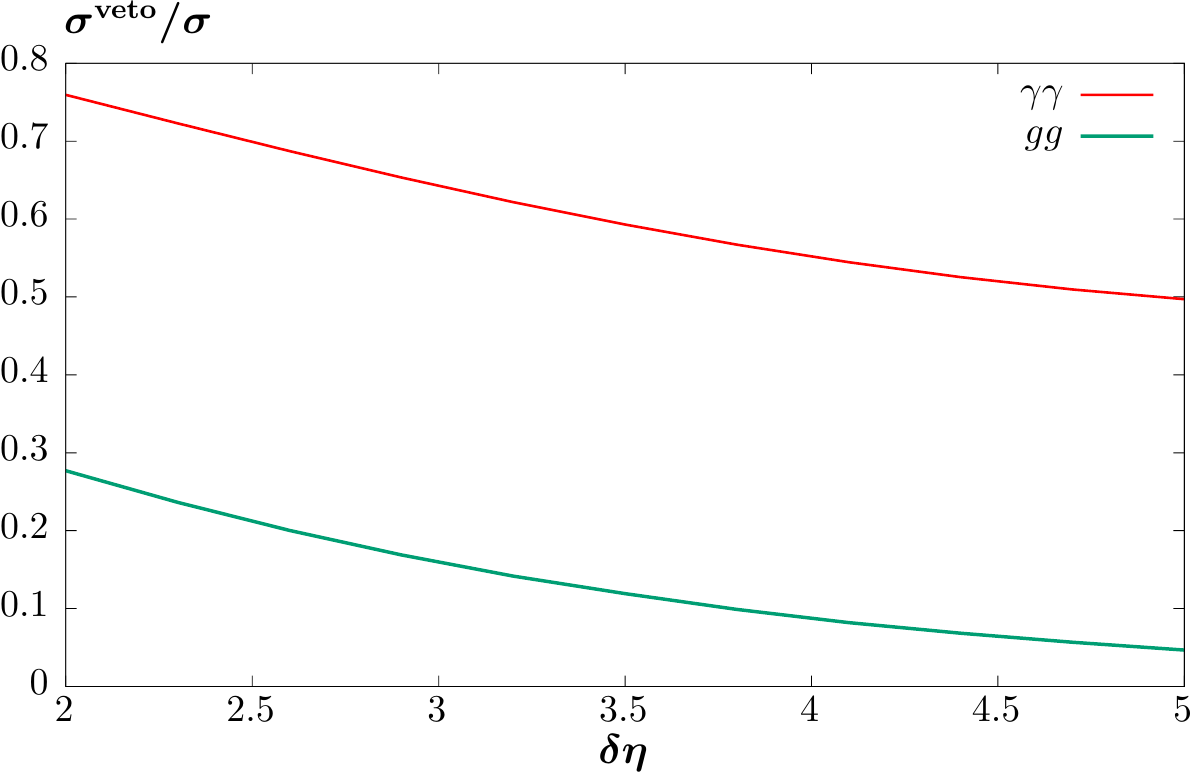}
\includegraphics[scale=0.65]{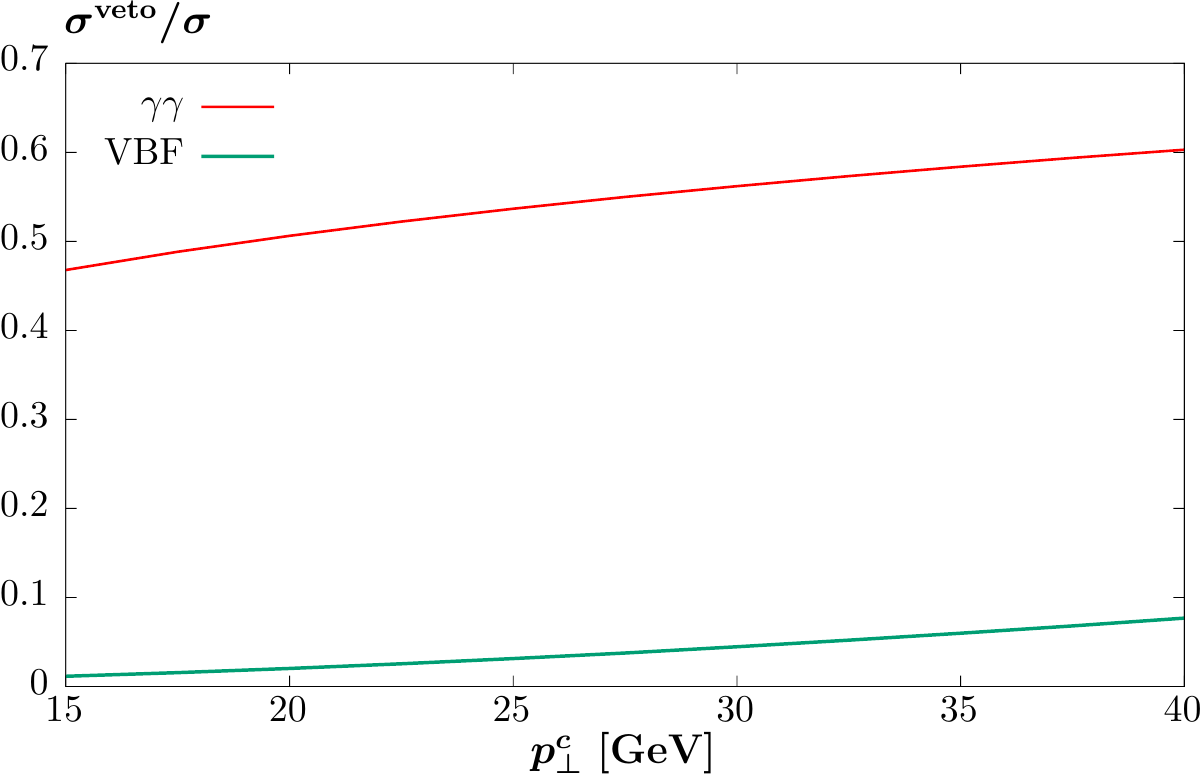}
\caption{Ratios of the $M_R=750$ GeV production cross section at $\sqrt{s}=13$ TeV subject to additional veto requirement to the inclusive cross section: (left) no extra emission in the interval $\delta\eta$ on either side of the resonance with transverse momentum $k_\perp>k_\perp^c=15$ GeV, for the $\gamma\gamma$ and $gg$ initiated processes, (right) the produced resonance is required to have transverse momentum $p_\perp^R<p_\perp^c$, for the $\gamma\gamma$ and VBF initiated processes.}\label{fig:epsg}
\end{center}
\end{figure}

Crucially, the effect of the veto (\ref{gg-supp}) on the $gg$ induced process will be much greater than in the $\gamma\gamma$ case. To first approximation, the $gg$ cross section will be  suppressed by a double logarithmic $S^2_g=e^{-2T_g}$ Sudakov factor with 
\be
T_g=\int_{{k_\perp^c}^{\!2}}^{\mu^2}\frac{\alpha_s(k^2_\perp)}{2\pi}\frac{dk^2_\perp}{k^2_\perp}
\int_0^{1-\Delta}\left[zP_{gg}(z)+\sum_q P_{qg}(z)\right]\Theta\left[\frac{k_\perp}{k_\perp+M_R e^{-\delta\eta}}-z\right]dz\;,
\label{sud-gg}
\ee
where $\Delta=k_\perp/(\mu+k_\perp)$, see~\cite{Kimber99} for more details, and here we have adjusted the conditions (\ref{gg-supp})  so that the virtual corrections corresponding to emission with momentum fraction $z$ inside the veto region are resummed. In Fig.~\ref{fig:epsg} (left) we show the ratio of the cross section with the veto applied and $k_\perp^c=15$ GeV, $\mu=M_R=750$ GeV, to the inclusive cross section for a range of $\delta\eta$ values and for the $\gamma\gamma$ and $gg$--initiated processes. While the $\gamma\gamma$--induced cross section is suppressed by just under a factor of $2$, the $gg$--induced cross section is further suppressed by $\sim 4-6$  for $\delta\eta=3-4$. Moreover, for the irreducible background from continuum diphoton production, which proceeds primarily through the $q\overline{q}$ channel, we can expect a similar level of supression. These analytical estimates therefore indicate that such a cut should give a fairly large suppression in the $gg$ contribution, although to be more precise a full MC simulation should be performed. We note this level of suppression is similar in size, although slightly smaller than, the result found in~\cite{Csaki:2016raa}, where instead tracks with transverse momentum above a very low threshold $p_\perp > 1$ GeV are vetoed on in a MC sample; such an approach leads to additional uncertainties due to the modelling of the underlying event and hadronization effects in the MC, and in addition such a stringent cut is found to reduce the `signal' $\gamma\gamma$--induced cross section by over a  factor of $\sim 10$.

In the case of VBF, considering $WW$ fusion for concreteness, the relevant part of the luminosity is proportional to
\be \label{lumiww}
\mathcal{L}_{WW}\propto \int\frac{{\rm d}^2 k_{1_\perp}}{(M_W^2+k_{1_\perp}^2)^2}\int\frac{{\rm d}^2 k_{2_\perp}}{(M_W^2+k_{2_\perp}^2)^2}
\;,
\ee
where $k_{i_\perp}$ are the transverse momenta transferred through the $W$ bosons, so that the resonance $\mathbf{p}_\perp^R=\mathbf{k}_{1_\perp}+\mathbf{k}_{2_\perp}$. In Fig.~\ref{fig:epsg} (right) we show the ratio of the cross section, subject to the requirement $p_\perp^R<p_\perp^c$, to the inclusive cross section, for a range of cut values\footnote{In fact, in the $\gamma\gamma$ case the transverse momenta cut is applied in the evolution of both PDFs, i.e. to the partons emitted in the last step of the evolution from both protons. This is not the same as applying the cut to the resonance transverse momenta, and indeed will overestimate the suppression somewhat, although for the relatively flat behaviour seen in Fig.~\ref{fig:epsg} (right) this is not a dramatic effect.}. The suppression in the VBF case is as expected from (\ref{lumiww}) of order $\sim (p_\perp^c)^2/M_W^2\ll 1$, and is very strong, while for the $\gamma\gamma$ luminosity we only expect a factor of $\sim 2$ reduction. Again,  although to be more precise a full MC simulation should be performed, these results indicate that a very strong reduction in any VBF contribution can be achieved with a simple cut choice.

\subsection{$\gamma\gamma$--initiated production: event structure}\label{sec:events}

In the previous section we have shown how the $\gamma\gamma$--initiated process is not dramatically reduced by requiring that no extra jets are present in a certain rapidity region surrounding the resonance. Here, we will comment a little further on the structure of the event that is expected for such a process, although for a more precise evaluation a MC study should be performed\footnote{We note that the results of~\cite{Csaki:2016raa} are consistent with the conclusions in this section. Moreover, since writing this paper we have confirmed these results with our own MC study, generating a 750 GeV scalar resonance + up to two jets with \texttt{MadGraph 5}~\cite{Alwall:2014hca} and matching this to parton shower generated with \texttt{Pythia} 8.215~\cite{Sjostrand:2006za,Sjostrand:2007gs}.}. To examine this further, we note that if we ignore the small corrections that the photon PDF will give to the evolution of the quark and gluons, then the equation (\ref{pdf}) for the DGLAP evolution of the photon PDF can be solved exactly, giving~\cite{Harland-Lang:2016apc}
\begin{align}\nonumber
\gamma(x,\mu^2)&=\gamma(x,Q_0^2)\,S_{\gamma}(Q_0^2,\mu^2)+\int_{Q_0^2}^{\mu^2}\frac{\alpha(Q^2)}{2\pi}\frac{dQ^2}
{Q^2}\int_x^1\frac{dz}z \bigg(\;\sum_q e^2_qP_{\gamma q}(z)q(\frac xz,Q^2)\\ \label{pdf1}
&+ P_{\gamma g}(z)g(\frac xz,Q^2)\bigg)\,S_{\gamma}(Q^2,\mu^2)\;,\\ \label{pdf1i}
&\equiv \gamma^{{\rm in}}(x,\mu^2)+\gamma^{\rm evol}(x,\mu^2)\;,
\end{align}
where the photon Sudakov factor
\be\label{sudgam}
S_{\gamma}(Q_0^2,\mu^2)=\exp\left(-\frac{1}{2}\int_{Q_0^2}^{\mu^2}\frac{{\rm d}Q^2}{Q^2}\frac{\alpha(Q^2)}{2\pi}\int_0^1 {\rm d} z\sum_{a=q,\,l}\,P_{a\gamma}(z)\right)\;,
\ee
corresponds to the probability for the photon PDF to evolve from scales $Q_0$ to $\mu$ without further branching; here $P_{q(l)\gamma}(z)$ is the $\gamma$ to quark (lepton) splitting function at NLO in $\alpha_s$. At LO it is given by
\be
P_{a\gamma}(z)=N_a\left[z+(1-z)^2\right]\;,
\ee
where $N_a=N_c e_q^2$ for quarks and $N_a=e_l^2$ for leptons, while the factor of $1/2$ in (\ref{sudgam}) is present to avoid double counting over the quark/anti--quarks (lepton/anti--leptons). The Sudakov factor is generated by resumming the term proportional to $P_{\gamma\gamma}$, due to virtual corrections to the photon propagator, which is a relatively small correction to the photon evolution. However for the reasonably large evolution length from $Q_0\sim$ 1 GeV to $\mu\sim 750$ GeV, this correction is not negligible, and we have $S_{\gamma}\sim 0.93$.

Thus, as shown in (\ref{pdf1i}) the photon PDF at a scale $\mu$ may be expressed as a sum of a term, $\gamma^{\rm in}(x,\mu^2)$, due to the input PDF, i.e. generated by coherent and incoherent photon emission up to the scale $Q_0$, multiplied by the probability of no further emission up to the hard scale $\mu$, and a second term, $\gamma^{\rm evol}(x,\mu^2)$, due purely to emission from the quark/gluons, which is independent of the input photon PDF.

These results, and those of the preceding section, allow us to make some relatively simple conclusions about the structure of a $\gamma\gamma$--initiated event. Firstly, upon inspection we find that the fraction of the photon PDF (\ref{pdf1i}) due to the input component is quite large: even for $\mu=750$ GeV it corresponds to $\sim 40\%$ of the total. This is generated either by coherent emission from the proton, or by incoherent emission from the individual quarks at low scale $\lesssim Q_0$, and will hence produce no secondary particles, up to soft underlying event activity. Thus from this fact alone, we expect that $\sim 16\%$ of events will be generated by this emission from both protons, and will therefore have no additional jet activity. Moreover, for the second `evolution' component due to quark/gluon DGLAP emission the transverse momenta of the emitted partons is often found to be fairly low: we can see from Fig.~\ref{fig:epsg} (right) that roughly $\sim 50\%$ of events are expected to have no additional jets with $k_\perp>20$ GeV.

We can also see from Fig.~\ref{fig:epsg} that about $\sim 65$\% of $\gamma\gamma$--initiated events are expected to have no additional jets with $k_\perp>15$ GeV up to $\pm 3$ units in rapidity from the resonance. For a higher cut, the fraction will of course be higher, although the dependence is not too strong: for e.g. $k_\perp>50$ GeV we expect $\sim 70$\% of events to have no jets in this rapidity interval. For a larger value of $\eta=\pm 5$, i.e. extending across essentially the entire ATLAS/CMS detector coverage, we expect $\sim 50(65)$\% of events to have no jets with $k_\perp>15(50)$ GeV\footnote{This implies that any invisible decay modes could be challenging to see, as any missing transverse energy will generally be small.}. Thus, by measuring the fraction of events with additional jets in these regions it should be possible to identify whether the resonance production mechanism is $\gamma\gamma$--initiated or not. Further information can also be provided by observing the fraction of events with jets on one side of the produced resonance: we recall from the discussion above that a sizeable fraction of the photon PDF from a given proton is generated by the low--scale input component, which will not produce any jets on the proton side. Finally, we note that the above conclusions are of a completely general nature, and would apply equally well to the production of other SM and BSM states via $\gamma\gamma$ fusion.

\section{Central Exclusive Production}\label{sec:CEP}

In addition to the inclusive channel considered above, for a resonance that is produced through $\gamma\gamma \to R$ it is natural to consider central exclusive production (CEP), $pp\to p+R+p$, where the protons remain intact after the collision. Such a final state is generated naturally by the colour--singlet $\gamma\gamma$ initial state; indeed, as discussed in Section~\ref{sec:gaminc} the dominant contribution to the photon PDF at the starting scale $Q_0$ is precisely from such coherent emission. We will evaluate below how this changes at the higher scale $\mu \sim M_R$ relevant to the resonance production process.

The exclusive channel is particularly relevant in light of the forward proton detectors approved for installation at ATLAS (AFP~\cite{CERN-LHCC-2011-012}) and already installed at CMS (CT-PPS~\cite{Albrow:1753795}): such exclusive events can be selected by tagging the outgoing intact protons in association with a measurement of the resonance $R$ in the central detector. The background from overlapping non--exclusive pile--up interactions may be controlled by ensuring that the `missing mass' and rapidity information reconstructed from the outgoing protons is consistent with the measurement in the central detector, as well as through the use of `fast timing' detectors to check if the photon and proton scattering points are the same, see~\cite{yp,Albrow:2011kt}.

By selecting exclusive events we naturally enhance the relative contribution from the $\gamma\gamma$--initiated subprocess, see~\cite{Khoze:2001xm}. In particular, for the $gg$--initiated case, which can occur exclusively through the `Durham' mechanism described in~\cite{Harland-Lang:2014lxa}, there is a strong Sudakov suppression (given by (\ref{sud-gg}) without the theta--function and with a much lower $k_\perp^c=Q_0=O({\rm GeV})$) associated with the requirement of no additional parton emission from the hard process. As a result, the exclusive $gg$ luminosity in the relevant kinematic regions is $\sim 3$ orders of magnitude smaller than in the inclusive case. In addition, for the final state to be exclusive there must be no underlying event activity associated with the hard process. The probability for this to occur is known as the `survival factor': see Appendix~\ref{sec:surv} for further discussion. For $gg$--induced production this suppresses the cross section by a further $\sim 2$ orders of magnitude, so that the exclusive cross section is suppressed in total by a very large factor of $\sim 10^5$.

In the $\gamma\gamma$--initiated process there is also some suppression from the fact that, while the dominant component of the input PDF, $\gamma(x,Q_0)$, is due to coherent emission from the proton, any further DGLAP evolution cannot occur, as this will produce secondary particles and spoil the exclusivity of the final state. More precisely, we calculate the exclusive $\gamma\gamma$ luminosity in the usual equivalent photon approximation (EPA)~\cite{Budnev:1974de}. The quasi--real photons are emitted by the incoming proton $i=1,2$ with a number density given by
\begin{equation}\label{WWflux}
n(x_i)=\frac{1}{x_i}\frac{\alpha}{\pi^2}\int\!\!\frac{{\rm d}^2q_{i_\perp} }{q_{i_\perp}^2+x_i^2 m_p^2}\left(\frac{q_{i_\perp}^2}{q_{i_\perp}^2+x_i^2 m_p^2}(1-x_i)F_E(Q_i^2)+\frac{x_i^2}{2}F_M(Q_i^2)\right)\;,
\end{equation}
where $x_i$  and $q_{i_\perp}$ are the longitudinal momentum fraction and transverse momentum of the photon $i$, respectively, and $Q_i^2$ is the modulus of the photon virtuality. The functions $F_E$ and $F_M$ are the usual proton electric and magnetic form factors
\begin{equation}\label{form1}
F_M(Q^2_i)=G_M^2(Q^2_i)\qquad F_E(Q^2_i)=\frac{4m_p^2 G_E^2(Q_i^2)+Q^2_i G_M^2(Q_i^2)}{4m_p^2+Q^2_i}\;,
\end{equation}
with
\begin{equation}\label{form2}
G_E^2(Q_i^2)=\frac{G_M^2(Q_i^2)}{7.78}=\frac{1}{\left(1+Q^2_i/0.71 {\rm GeV}^2\right)^4}\;,
\end{equation}
in the dipole approximation, where $G_E$ and $G_M$ are the `Sachs' form factors. The `EPA' $\gamma\gamma$ luminosity is  given by
\begin{equation}\label{lgam}
\frac{{\rm d}\mathcal{L}^{\rm EPA}_{\gamma\gamma}}{{\rm d}M_X^2\,{\rm d}y_X}= \frac{1}{s}\,n(x_1) \,n(x_2)\;.
\end{equation}
Comparing with (\ref{lumiinc}) we can see that the number density $n(x_i)$ corresponds to the coherent component of the photon PDF. To calculate the contribution from this coherent component at the scale $\mu\sim$ 750 GeV we must then multiply this by the Sudakov factor (\ref{sudgam}), corresponding to the probability that the coherently emitted photon does not split into quarks or leptons, spoiling the exclusivity of the final state. As discussed in Section~\ref{sec:events}, this leads to a suppression of $S^2_\gamma \sim 0.86$: while therefore much less significant than in the QCD--initiated case, where the probability of additional branching for the initial--state gluons is much higher, this can nonetheless not be ignored entirely. Comparing to the inclusive luminosity, we find that requiring exclusivity in the DGLAP evolution reduces the cross section by roughly an order of magnitude.

\begin{figure}
\begin{center}
\includegraphics[scale=0.65]{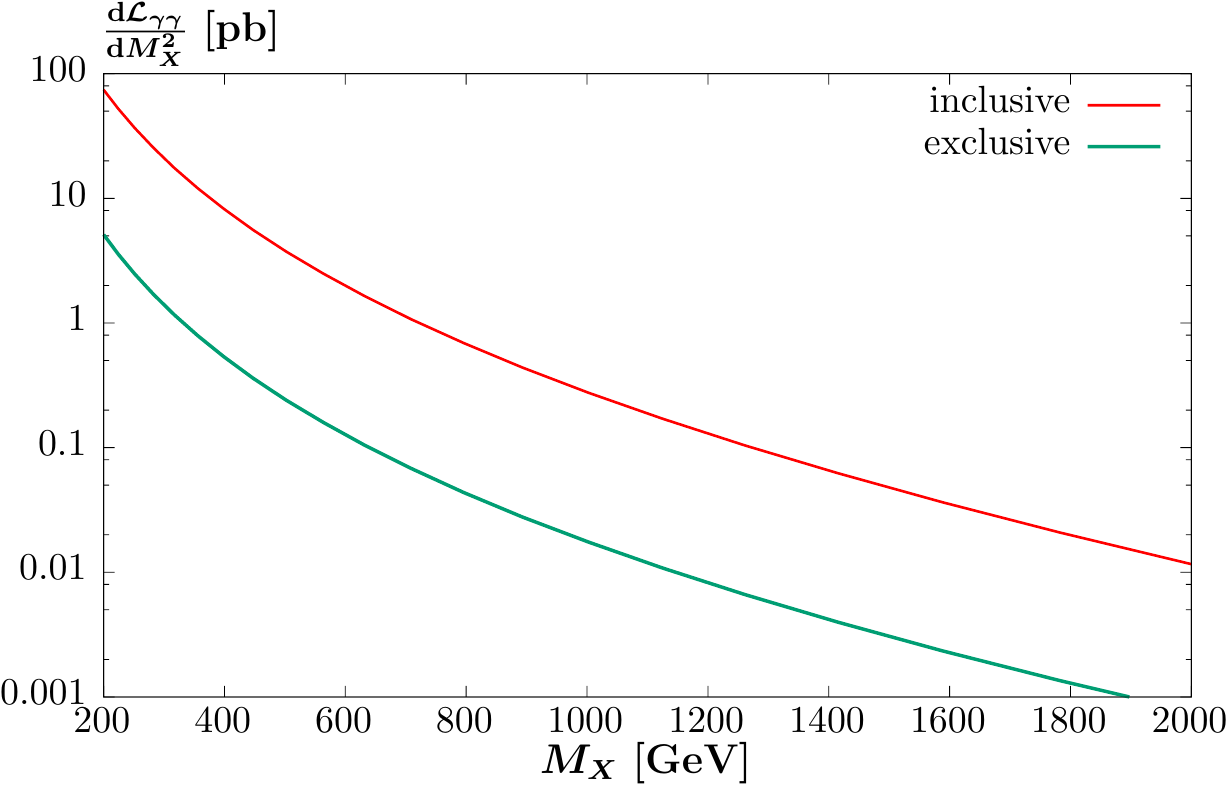}
\includegraphics[scale=0.65]{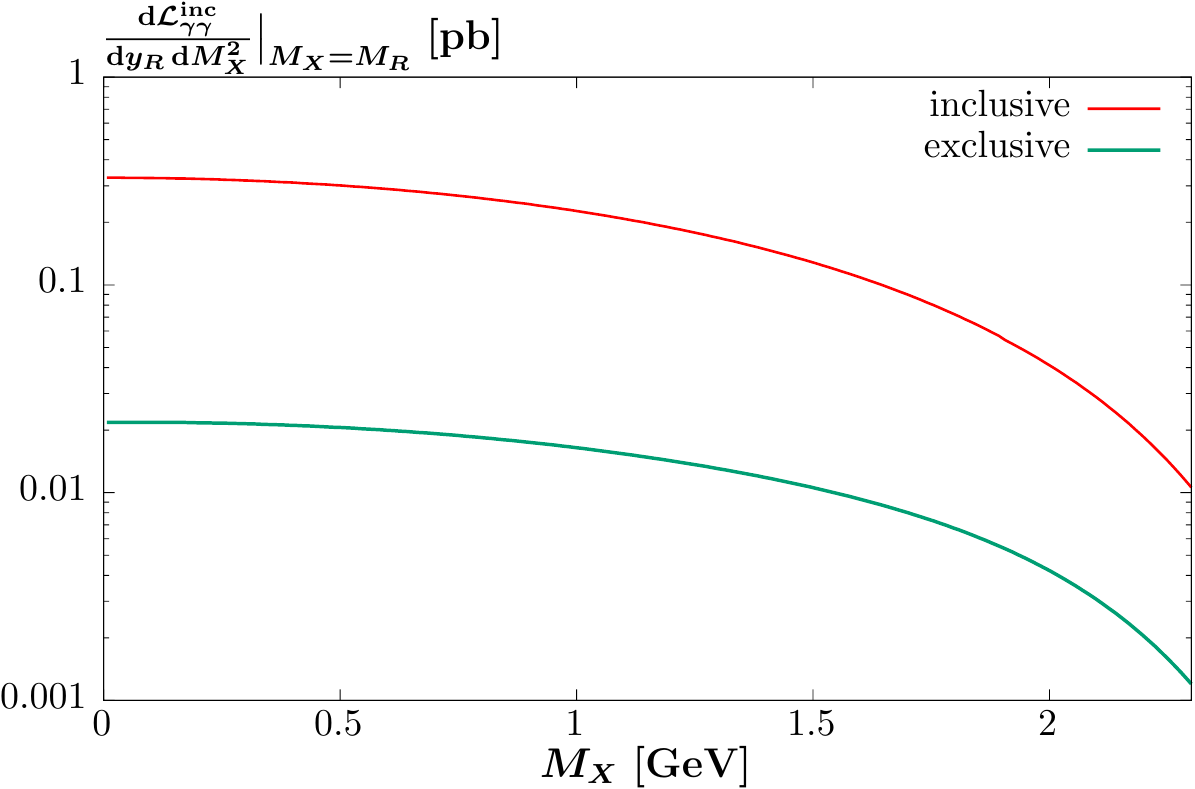}
\caption{Inclusive and exclusive $\gamma\gamma$ luminosities for a scalar resonance at $\sqrt{s}=13$ TeV, shown (left) differential in the invariant mass, $M_X$, of the produced system, integrated over rapidity $y_X$, (right) for the production of a resonance of mass $M_R=750$ GeV, differential in the particle rapidity.}\label{fig:lumiexc}
\end{center}
\end{figure}

However, the formulae presented above are in fact only approximately correct for the case of exclusive production; we must in addition consider the effect of the survival factor, as above. In such a situation we cannot na\"{i}vely apply (\ref{WWflux}), but rather we must correctly account for the polarization structure of the $\gamma\gamma\to R$ process at the amplitude level, as described in detail in Appendix~\ref{sec:surv}. The coherently emitted photons have relatively low transverse momentum, $q_\perp$, corresponding to larger impact parameters between the colliding protons, where the probability of additional particle production is small. Thus, the suppression is in fact not too great, with the precise value depending on the parity of the produced resonance: we get $S^2= 0.72$(0.77) for a 750 GeV scalar(pseudoscalar) state. In Fig.~\ref{fig:lumiexc} we show the exclusive and inclusive $\gamma\gamma$ ratios for scalar resonance production as function of the invariant mass $M_X$ of the produced state, and for the production of a resonance of mass $M_R=750$ GeV, as a function of rapidity. In the latter case, we see that a factor of $\sim 16$ reduction is induced in the luminosity by requiring exclusivity, with more precisely
\begin{equation}\label{sigexc}
\sigma^{\rm exc}(pp \to (R\to\gamma\gamma))=0.063\cdot\sigma^{\rm inc}(pp \to (R\to\gamma\gamma))= 0.25 - 0.50\,{\rm fb}\;,
\end{equation}
for inclusive cross sections in the range $4-8$ fb. 

What is the uncertainty on this? As discussed in Section~\ref{sec:gaminc}, we expect a $\sim 15-20\%$ uncertainty in the inclusive $\gamma\gamma$ luminosity, due principally to the factorization scale variation. However, in the purely exclusive case the prediction is under even better theoretical control. The initial--state is no longer given in terms of inclusive photon PDFs with corresponding factorization scale uncertainty, associated with higher order QCD corrections; rather we must only consider the probability for the entire proton to coherently produce a photon. This is very well understood, with the coupling of the coherent photon to the proton parameterised by the experimentally well--constrained form factors (\ref{form1}) and (\ref{form2}). There is some uncertainty due to the choice of scale in the Sudakov factor (\ref{sudgam}), but this is small: varying $\mu$ between $M_R/2$ and $M_R$ we get a $\sim \pm 2\%$ variation in $S_\gamma^2$. Another question relates to the survival factor. However, as discussed above, for photon--induced processes, the average impact parameter between the colliding protons is generally large. This is in a regime where the proton optical density $\Omega(b_t)$, which is required to calculate the survival factor  (see Appendix~\ref{sec:surv}), is well constrained to be small in size. This means that any allowed variation in its value  results in a very small change in the probability of no inelastic production $\exp(-\Omega(b_t))$; in~\cite{Harland-Lang:2015cta} various models for $\Omega(b_t)$ are taken for a range of photon--induced processes, and the variation in the resulting cross sections is found to be  at the $\sim \%$ level. Thus in this exclusive case the expected theoretical uncertainty is extremely small, of order a few percent.

Finally, it is worth pointing  out that the resolution of the missing mass measured by tagging the outgoing protons is expected in this region to be rather good, with $\Delta M\sim 10$ GeV~\cite{yp}). Thus, if the relatively high width suggested by the ATLAS data is in fact due to a superposition of more than one resonances of similar masses (see for example~\cite{Djouadi:2016eyy}), the exclusive mode could allow these to be separated.  A further possibility is that if the resonance has a sizeable decay to invisible particles (e.g. dark matter~\cite{Abel:2016pyc}), then in general this may be observed in the `missing mass' spectrum reconstructed from the tagged outgoing protons. However, in light of backgrounds from e.g. pile--up interactions and low mass diffractive dissociation or elastic scatters combined with photon emission from protons (see~\cite{Belotsky:2004ex} for more details) this appears to be an extremely challenging measurement at the required luminosities.

\subsection{Spin-parity analysis}\label{sec:spin}

As well as being sensitive to the $\gamma\gamma$--induced production mechanism, the exclusive mode offers the additional advantage that by tagging the outgoing protons, further information may be provided about the quantum numbers of the produced state. This was discussed in~\cite{Kaidalov:2003fw} for the $gg$--mediated process, but a similar situation applies here.

We will consider as an example the possibility to distinguish between a scalar and a pseudoscalar resonance. Differentiating between these two possibilities from measurements of the $\gamma\gamma$ final--state in the central detector is not possible, as the photon angular distributions are identical in the scalar and pseudoscalar cases. However, in the exclusive mode the situation is much more encouraging: by measuring the outgoing proton transverse momenta we are in fact directly sensitive to the polarisation structure of the $\gamma\gamma \to R$ process, and hence to the quantum numbers of the resonance. In particular, in exclusive interactions the $\gamma\gamma \to R$ subprocess amplitude can be written as
\be
A\sim q_{1_\perp}^\mu q_{2_\perp}^\nu V_{\mu\nu}\;,
\ee
where $V_{\mu\nu}$ is the usual $\gamma\gamma \to R$ vertex, see~\cite{Harland-Lang:2015cta} for further discussion, and $q_{i_\perp}$ are the photon transverse momenta. Comparing to the coupling of $R$ to external photons, we can see that the $q_{i_\perp}$ play the role of the photon polarisation vectors, which are therefore directed along their transverse momenta. 

\begin{figure}
\begin{center}
\includegraphics[scale=0.65]{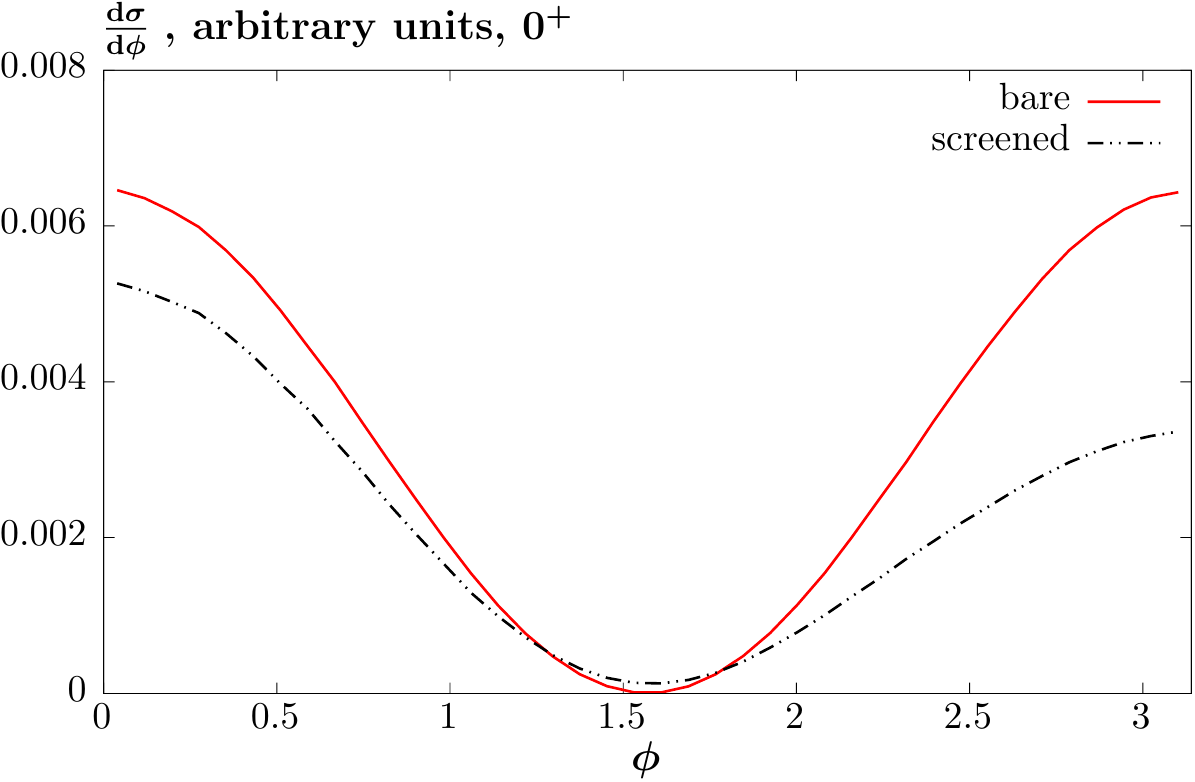}
\includegraphics[scale=0.65]{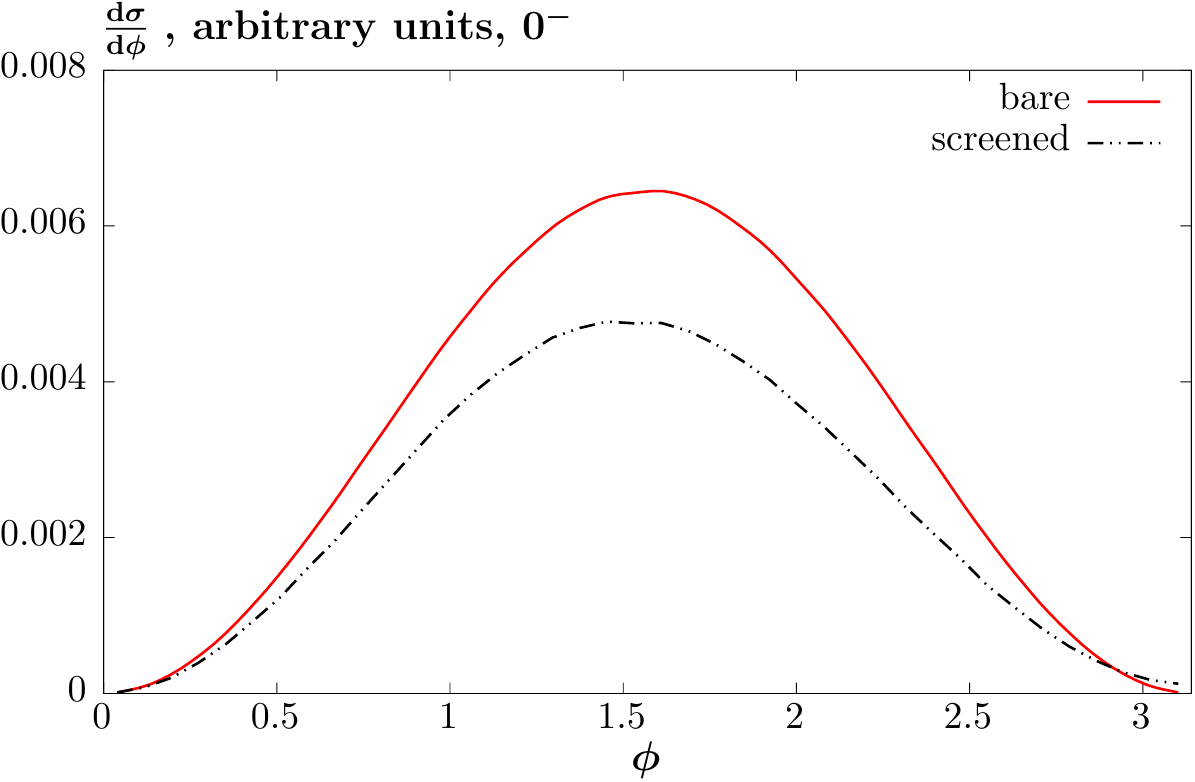}
\caption{Distribution, in arbitrary units, with respect to the azimuthal angle $\phi$ between the transverse momenta of the outgoing protons for the exclusive production of a scalar (left) and pseudoscalar (right) resonance of mass $M_R=750$ GeV at $\sqrt{s}=13$ TeV.}\label{fig:cor}
\end{center}
\end{figure}

Crucially, in this exclusive case any transverse momentum transferred through the incoming photons must be compensated by a corresponding transverse momentum, $p_\perp$, of the outgoing protons. Indeed, to good approximation (this ignores the influence of the survival factor, which we will discus below) the photon transverse momentum is simply anti--aligned with the recoil photon, i.e. we have $p_{i_\perp}=-q_{i_\perp}$, where $i=1,2$. The polarisation structure of the amplitudes in the scalar and pseudoscalar cases have quite distinct forms and, writing in terms of the outgoing proton transverse momenta we have 
\begin{align}\label{ap}
|A^+|^2 &\sim |p_{1_\perp}\cdot p_{2_\perp}|^2\sim \cos^2 \phi\;,\\ \label{am}
|A^-|^2 &\sim |\epsilon_{\alpha\beta\mu\nu}\,p_1^\alpha p_2^\beta p^{\mu}_{1_\perp} p^{\nu}_{2_\perp}|^2\sim \sin^2\phi\;,
\end{align}
where the $+(-)$ indicate the scalar(pseudoscalar) cases, $p_i$ is the 4--momentum of proton $i$, and $\phi$ is the azimuthal angle between the outgoing proton $p_\perp$ vectors. Thus, we expect quite distinct azimuthal correlations between the outgoing protons for the two cases, which can readily be measured by tagging detectors.

More precisely, these simple $\cos^2(\phi)$ or $\sin^2(\phi)$ distributions are in fact somewhat distorted by the  influence of `absorptive' corrections which generate the survival factor, $S^2$, i.e. the probability that additional particles are not produced as a result of soft proton--proton interactions. As discussed in more detail in Appendix~\ref{sec:surv}, to account for survival effects we must include an additional elastic interaction between the protons, with momentum transfer $k_\perp$. This will then be transferred through the photon propagators as well, so that we no longer have the exact relation $p_{i_\perp}=-q_{i_\perp}$ between the photon and outgoing proton transverse momenta. Nonetheless, the average momentum transfer is small, with $k_\perp^2 \sim 2/ B_{\rm el}$, where the $t$--slope for elastic $pp$ scattering $B_{\rm el}\sim 20\,{\rm GeV}^{-2}$ at the LHC~\cite{Antchev:2013gaa,Aad:2014dca}, and so $k_\perp^2\sim 0.1 \, {\rm GeV}$. Thus, after integrating over $k_\perp$, the exact $\cos^2(\phi)$ or $\sin^2(\phi)$ distributions are washed out somewhat, but the dominant behaviour remains.

This is seen in Fig.~\ref{fig:cor}, where we plot the predicted $\phi$ distribution for scalar (left) and pseudoscalar (right) particles. These are calculated using a modified version of the \texttt{SuperChic 2} MC~\cite{Harland-Lang:2015cta}, with model 4 of~\cite{Khoze:2013dha} taken for the survival factor (the resulting distributions are not sensitive to this precise choice). The difference between the scalar and pseudoscalar cases is clear, with the distinct azimuthal distributions resulting from (\ref{ap}) and (\ref{am}) remaining even after survival effects are included. With only a few observed signal events, it could be possible to distinguish between these two scenarios, due to the dramatically different behaviour predicted at $\phi=0$ and $\pi$.

\subsection{CP--violating effects}\label{sec:cp}
 
 As well as being sensitive to the parity of the produced object, any $CP$--violation in the production mechanism of the resonance $R$ will in fact induce an asymmetry in the proton $\phi$ distribution. This was shown in~\cite{Khoze:2004rc} in the context of light $CP$--violating Higgs production, for which the $gg$--induced production amplitude is given by
\be
A^{CPV}\sim g_s\, p_{1_\perp}\cdot p_{2_\perp} - \frac{g_P}{p_1\cdot p_2} \, \epsilon_{\alpha\beta\mu\nu}\,p_1^\alpha p_2^\beta p^{\mu}_{1_\perp} p^{\nu}_{2_\perp}\;,
\ee
where $g_S$ and $g_P$ are the corresponding couplings. In the case of exclusive $\gamma\gamma$--initiated production we expect the same form of amplitude. As the first term has a $\sim \cos\phi$ behaviour, while the second instead is $\sim \sin\phi$, it can readily be shown that upon squaring the amplitude, the interference between these two 
 terms leads to an asymmetry in the predicted proton $\phi$ distribution. Thus, a measure of the asymmetry
 \be
 \mathcal{A}=\frac{\sigma(\phi<\pi)-\sigma(\phi>\pi)}{\sigma(\phi<\pi)+\sigma(\phi>\pi)}
 \ee
 will be sensitive to $CP$--violating effects-- see~\cite{Khoze:2004rc} for a detailed discussion.

\subsection{CEP in heavy ion collisions}\label{sec:ion}

At first sight, an attractive possibility is  to study the photon--induced CEP of a resonance at $750$ GeV  in heavy ion collisions, where the coherent photon flux is generally enhanced by the squared charge of the beam, $Z^2$, from both sides, see e.g.~\cite{d'Enterria:2013yra}. However, in the kinematic regime corresponding to the production of such an object, where the photon $x$ is quite high, the situation is not encouraging. In particular, the minimum squared photon virtuality  $t_{min}\simeq -(xm_p)^2/(1-x)$ $\sim (140\, {\rm MeV})^2$ for a typical $\sqrt{s_{NN}}=5.5$ TeV for Pb--Pb collisions, corresponding to a photon `transverse size' of $\sim 1.4$ fm. As this number is significantly smaller than the radius of the ion, e.g. we have $R_A\sim 7$ fm for lead, the possibility for coherent emission from the entire heavy ion nucleus is greatly reduced. Indeed, taking the standard form for the $\gamma$ flux from a lead ion, as in e.g.~\cite{Baltz:2007kq}, the expected cross section in Pb--Pb is similar in size, $\sim$ fb, to the proton--proton case; the $Z^2$ enhancement is essentially lost. Thus, for the much lower luminosities that can be expected in heavy ion runs at the LHC, such a measurement appears to be unrealistic.

\section{Production of colourless fermion pairs}\label{sec:fermion}

In many models the decay of the diphoton resonance is mediated via an intermediate loop formed of sleptons~\cite{Allanach:2015ixl,Ding:2015rxx}, vector quark or leptons~\cite{Hall:2015xds} or fermions~\cite{Fichet:2015vvy}. For illustration we will consider the fermion case in what follows, although similar results may be found for other particle types. In principle, if such fermions do not couple to colour they may be within the mass reach of the LHC, but have not yet been observed; indeed, as discussed in e.g.~\cite{Djouadi:2016eyy}, the existence of such states is still relatively unconstrained for masses above $\sim 200$ GeV.

 Ignoring QED threshold effects, the $\gamma\gamma\to F\bar F$ subprocess cross section is given by (see e.g.~\cite{Isparin:1970ye})
\be
\frac{{\rm d}\hat{\sigma}}{{\rm d}\cos\theta^*}(\gamma\gamma\to F \overline{F}) = e_F^4\frac{2\pi\alpha^2\beta}{M_X^2}\frac{1+2\beta^2(1-\beta^2)(1-\cos^2\theta^*)-\beta^4\cos^4\theta^*}{(1-\beta^2\cos^2\theta^*)^2}\,,
\ee
where $\beta=(1-4m_F^2/M_X^2)^{1/2}$, $e_F$ is fermion electric charge, $m_F$ is the fermion mass, and $M_X$ is the $F\overline{F}$ invariant mass. Combining this with the inclusive $\gamma\gamma$ luminosity given in the preceding sections, and for illustration taking $m_F=360$ GeV and $e_F=1$, we get $\sigma_{F\overline{F}}=0.12$ fb at $\sqrt{s}=13$ TeV. While this is quite small, in particular relative to the suggested resonance $R$ production cross section, this will be strongly enhanced in a scenario where the new fermion carry higher electric charge $e_F>1$. Note that the resonant $R\to F\overline{F}$ cross section may give a comparable contribution to the overall $F\overline{F}$ signal, provided the corresponding branching ratio is not too small.

\section{Conclusion}\label{sec:conc}

The observation by the ATLAS and CMS collaborations of an excess of events around 750 GeV in the diphoton mass spectrum has recently provoked a great deal of theoretical interest. As the only hint of any discrepancy from the SM in this mass region is so far in the $\gamma\gamma$ channel (and not e.g. in the dijet mass spectrum), an essentially minimal interpretation of the signature is that it is due to the decay of a resonance $R$ which couples only, or at least dominantly, to photons. In such a scenario the main production mechanism, as well as the decay channel, will be $\gamma\gamma$--mediated.

In this paper, we have considered the case of a scalar or pseudoscalar resonance $R$ of mass 750 GeV, which is produced through $\gamma\gamma$ collisions. This may occur inclusively or exclusively, with in the latter case the outgoing protons remaining intact after the interaction. Our aim has not been to present results within the context of a particular model, but rather to provide the most precise possible predictions for the $\gamma\gamma$ luminosity, needed to calculate the corresponding resonance production cross sections, in both the inclusive and exclusive cases. 

The precise numbers we have presented (which depend on the resonance mass, width and branching ratios) are for illustration only, as the available experimental information is currently quite limited. Nonetheless, these predictions, and the discussion we present here, indicate how any future analysis can be performed, if after gathering more data the excess remains, and the properties of the underlying resonance become clearer. Moreover, the calculations presented in this paper are not only applicable to the case of such a resonance: the $\gamma\gamma$--initiated channel, both exclusive and inclusive, is potentially sensitive to a range of BSM physics, see e.g.~\cite{deFavereaudeJeneret:2009db,Fichet:2014uka,yp}.

The main results of this paper are as follows:

\begin{itemize}

\item The inclusive $\gamma\gamma$ luminosity has been calculated with NLO accuracy, with an uncertainty of $\sim 15-20$ \%, principally associated with the choice of factorization scale and input photon PDF. 

\item The ratio of inclusive cross sections for a 750 GeV resonance at $13$ to $8$ TeV is found to be $\sim 2.9$. This result is consistent with the CT14~\cite{Schmidt:2015zda} and MRST2004~\cite{Martin:2004dh} QED PDF sets, which include some theoretical guidance for the form of the photon PDF. Although the NNPDF2.3QED~\cite{Ball:2013hta} set, which takes a completely free parameterisation of the  photon PDF, prefers a lower central value $\sim 2.1$, this is in a region where the corresponding photon PDF in this approach is relatively unconstrained, and the uncertainty on the ratio is quite large.

\item Simple cuts on the final state can efficiently reduce the relative contribution from the $gg$ and VBF production channels, if such modes are present, relative to the $\gamma\gamma$--initiated case.

\item A precise calculation of the exclusive $\gamma\gamma$ luminosity, relevant to the case where both protons remain intact after the interaction, has been presented, with an associated uncertainty that is very small, and does not exceed a few percent. 

\item  The exclusive channel leads naturally to a strong suppression of the $gg$ and VBF initiated modes. The ratio of inclusive to exclusive $\gamma\gamma$ luminosities is found to be $\sim 16$, with corresponding exclusive cross section $\sim 0.3-0.5$ fb via the $\gamma\gamma$ decay channel, for the current best estimate of the inclusive cross section corresponding to the apparent diphoton excess. Assuming favourable experimental efficiencies and resolution this could therefore be accessible with the hundreds of ${\rm fb}^{-1}$ of integrated luminosity which can be taken with the AFP~\cite{CERN-LHCC-2011-012,Tasevsky:2015xya} and CT--PPS~\cite{Albrow:1753795} forward proton taggers, associated with the ATLAS and CMS central detectors, respectively. It is in particular worth pointing out that the mass of the potential resonance is precisely in the region of maximum acceptance for these detectors~\cite{yp}. 

\item The exclusive channel allows the outgoing intact protons to be measured by tagging detectors. We have demonstrated that the predicted distribution with respect to the azimuthal angle between the proton $p_\perp$ vectors is highly sensitive to the parity of the produced object, and that with just a handful of events the scalar and pseudoscalar cases may be distinguished.

\item The \texttt{SuperChic 2} MC~\cite{Harland-Lang:2015cta}, gives a state of the art and precise treatment of a range of exclusive $\gamma\gamma$--initiated processes, including soft survival effects.

\item Although na\"{i}vely one might assume that heavy ion collisions are a natural place to look for the production of a 750 GeV resonance which couples dominantly to photons, we have shown that the $Z^2$ enhancement is essentially lost in the kinematic regime relevant to such a heavy object, due to the relatively high average photon virtuality. Consequently, the predicted rates are too low for such an observation to be realistic.

\end{itemize}

It remains entirely possible that the excess of events observed by ATLAS and CMS is a purely statistical fluctuation. However, if after collecting more data the signal remains, it is the task of theorists to provide the most precise and up--to--date possible predictions for the expected experimental signatures. It has been our aim in this paper to achieve this for the case that the resonance persists and couples dominantly to photons; if this is the case, then a variety of interesting studies, in both the exclusive and inclusive channels, are possible, and we hope to lay the groundwork for these here.

.

\section*{Acknowledgements}

We thank Michael Spannowsky for invaluable advice on MC generation and for other useful discussion. We also thank Steve Abel, Valya Khoze, and Marek Tasevsky for useful discussions. VAK thanks the Leverhulme Trust for an Emeritus Fellowship. The work of MGR  was supported by the RSCF grant 14-22-00281. LHL thanks the Science and Technology Facilities Council (STFC) for support via the grant award ST/L000377/1. MGR thanks the 
IPPP at the University of Durham for hospitality.

\appendix

\section{Soft survival factor}\label{sec:surv}

The survival factor, denoted $S^2$, corresponds to the probability of no additional underlying event activity, i.e. additional soft particle production\footnote{In this paper we only consider the so--called `eikonal' survival factor, due to additional proton--proton interactions. In general, we should also consider the so--called `enhanced' survival factor, see e.g.~\cite{Ryskin:2011qe}, generated by additional interactions with the intermediate partons produced during the evolution. However for exclusive $\gamma\gamma$--initiated processes this is absent.}.  It is crucial to include this when calculating any exclusive cross section, where we require that the protons remain intact and there is no hadronic production  in addition to the considered final--state (in this paper the decay products of the resonance $R$); the underlying event will clearly spoil this exclusivity requirement. As the survival factor is a soft physics, and hence non--perturbative, object it cannot be calculated from first principles, and a phenomenological model must be used. Typically a `global' approach is taken, and soft QCD models which predict a range of hadronic observables, such as the the total, elastic and diffractive cross sections, as well as the survival factor, are in fact quite well developed (see e.g.~\cite{Khoze:2013dha,Khoze:2014aca,Gotsman:2014pwa} for recent studies). These can therefore be tuned to such data, allowing the size of $S^2$ to be fairly well constrained.

\begin{figure}
\begin{center}
\includegraphics[scale=0.25]{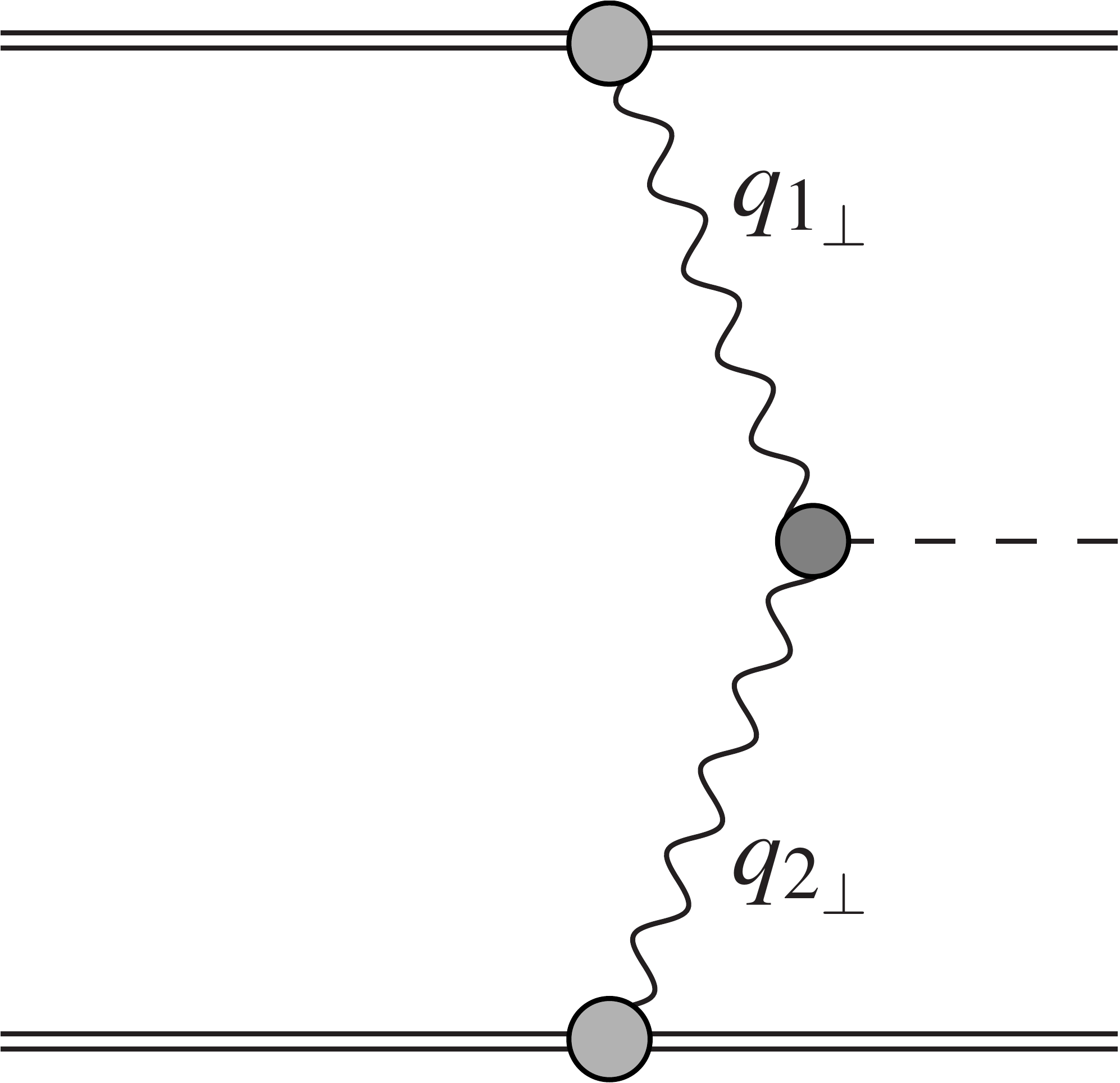}\qquad\qquad
\includegraphics[scale=0.3]{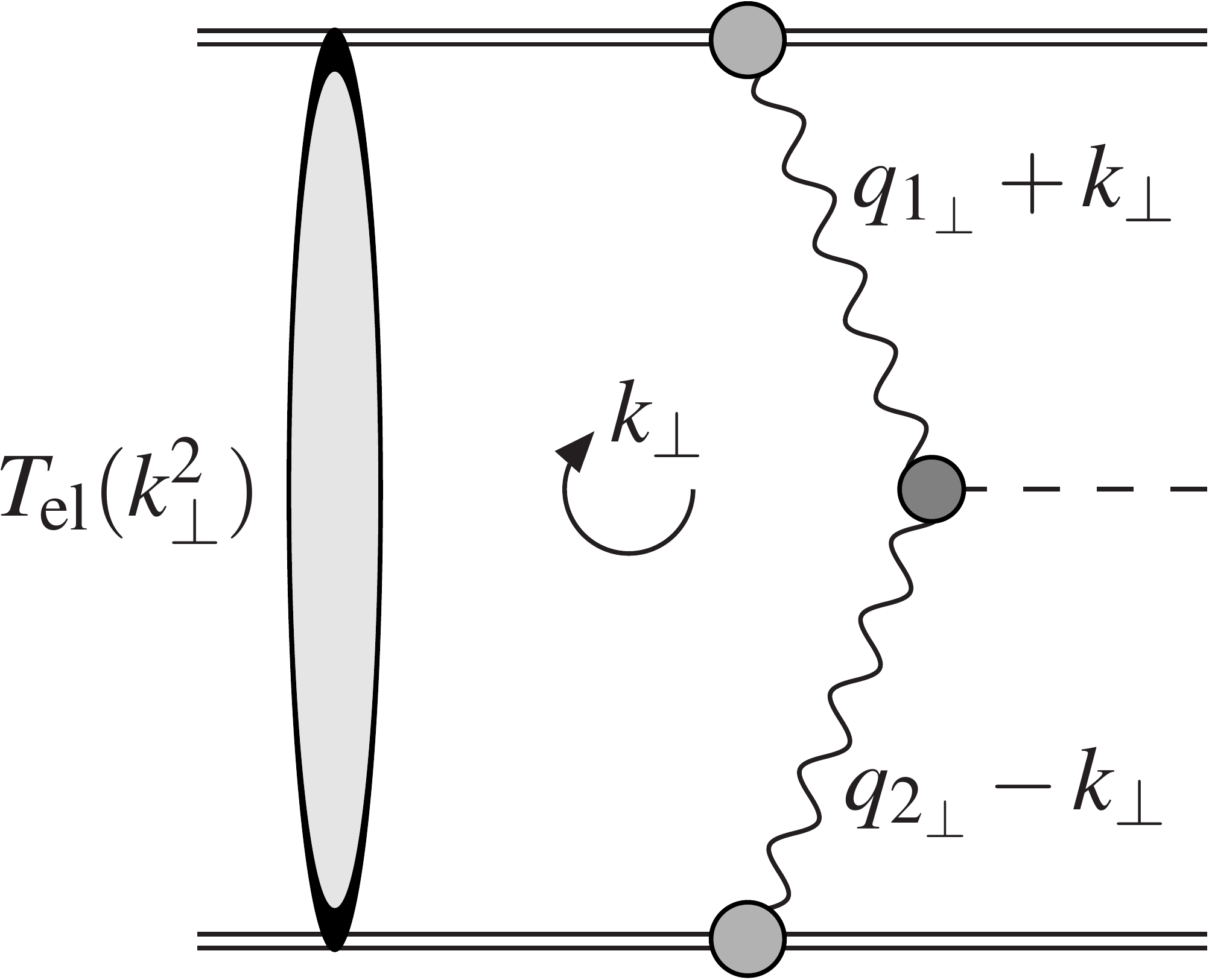}
\caption{Feynman diagrams for (left) bare and (right) screened amplitudes for exclusive  $\gamma\gamma$--initiated resonance production.}\label{fig:scr}
\end{center}
\end{figure}

One point to emphasise is that the survival factor is not a simple multiplicative constant~\cite{HarlandLang:2010ep}, but rather depends in general on the final--state configuration, and in particular on the outgoing proton transverse momenta. Physically, this is to be expected, as the survival factor will  depend on the impact parameter of the colliding protons; loosely speaking, as the protons become more separated in impact parameter, we should expect there to be less additional particle production, and so for the survival factor to be closer to unity. As the transverse momenta ${\bf p}_{i_\perp}$ of the scattered protons are nothing other than the Fourier conjugates of the proton impact parameters, ${\bf b}_{it}$, we therefore expect the survival factor to depend on these. It is precisely this effect which leads the survival factor for the relatively peripheral $\gamma\gamma$--initiated processes to be close to unity, as discussed in Section~\ref{sec:CEP}.

The above considerations therefore imply that the survival factor should be treated differentially: this was achieved within the \texttt{SuperChic 2} MC framework in~\cite{Harland-Lang:2015cta}, and we give a brief summary of how this is done for $\gamma\gamma$--initiated processes below. The diagram in Fig.~\ref{fig:scr} (left) corresponds to the usual so--called `bare' amplitude, prior to including any survival effects. Taking a scalar resonance $R$ for illustration, and recalling (\ref{ap}) and (\ref{WWflux}) we can write this as
\begin{equation}\label{tbare}
T(q_{1_\perp},q_{2\perp}) \sim \frac{F_E(Q_1^2)^{1/2}}{q_{1_\perp}^2+\xi_1^2 m_p^2}\, \frac{F_E(Q_2^2)^{1/2}}{q_{2_\perp}^2+\xi_2^2 m_p^2}\,(q_{1_\perp}\cdot q_{2_\perp}) \;,
\end{equation}
where we do not show the contribution from the magnetic form factor for simplicity (see~\cite{Harland-Lang:2015cta} for a discussion of how this can be included), and overall factors due to the $\gamma\gamma \to R$ vertex and $x_i$ dependence (and other factors) from the photon flux (\ref{WWflux}) are omitted for clarity. In this case the transverse momenta $q_{i_\perp}$ transferred through the photons must be exactly balanced by the outgoing proton momenta, and so we have $q_{i_\perp}=-p_{i_\perp}$ and $T(q_{1_\perp},q_{2\perp})=T(p_{1_\perp},p_{2\perp}) $.

It can be shown~\cite{Gribov:1968fc} that to calculate the survival probability we must simply consider the additional diagram show in Fig.~\ref{fig:scr} (right), where the grey oval represents an additional proton--proton elastic scatter, where a transverse momentum $k_\perp$ is exchanged, and with corresponding amplitude $T^{\rm el}(k^2_\perp)$.  For this `screened' amplitude we must integrate over the momentum $k_\perp$ transferred through the loop, and so we have
\be\label{tscr}
T^{\rm scr.}(p_{1_\perp},p_{2_\perp})=\frac{i}{s}\int\frac{{\rm d}^2k_\perp}{8\pi^2}\,T^{\rm el}(k^2_\perp)\,T(q'_{1_\perp},q'_{2_\perp})\;,
\ee
where, as the $k_\perp$ exchanged by the elastic scatter is transferred through the photon propagators, we have $q'_{1_\perp}=-p_{1_\perp}+k_\perp$ and $q'_{2_\perp}=-p_{2_\perp}-k_\perp$. We must then add this to the bare amplitude to give the final result for the differential cross section
\be
\frac{{\rm d} \sigma}{{\rm d}p_{1_\perp}{\rm d}p_{2_\perp}}\sim |T(p_{1_\perp},p_{2_\perp})+T^{\rm scr.}(p_{1_\perp},p_{2_\perp})|^2\;.
\ee
As the elastic amplitude, $T^{\rm scr.}$ is dominantly imaginary, from (\ref{tscr}) we can see that this interferes destructively with the bare amplitude, reducing the cross section. The average survival factor is simply
\be
S^2  = \frac{\int {\rm d}^2 p_{1_\perp}{\rm d}^2 p_{2_\perp} |T(p_{1_\perp},p_{2_\perp})+T^{\rm scr.}(p_{1_\perp},p_{2_\perp})|^2}{\int {\rm d}^2 p_{1_\perp}{\rm d}^2 p_{2_\perp}  |T(p_{1_\perp},p_{2_\perp})|^2}\;,
\ee
which is by construction always less than one. In particular, it can shown that in impact parameter space this is equivalent to 
\begin{equation}\label{S2}
S^2 =\frac{\int {\rm d}^2 {\bf b}_{1t}\,{\rm d}^2 {\bf b}_{2t}\, |T(s,{\bf b}_{1t},{\bf b}_{2t})|^2\,{\rm exp}(-\Omega(s,b_t))}{\int {\rm d}^2\, {\bf b}_{1t}{\rm d}^2 {\bf b}
_{2t}\, |T(s,{\bf b}_{1t},{\bf b}_{2t})|^2}\;,
\end{equation}
where ${\bf b}_{it}$ is the impact parameter vector of proton $i$, so that ${\bf b}_t={\bf b}_{1t}+{\bf b}_{2t}$ corresponds to the transverse separation between the colliding protons, with $b_t = |{\bf b}_t|$.  $T(s,{\bf b}_{1t},{\bf b}_{2t})$ is the amplitude (\ref {tbare}) in impact parameter space, and $\Omega(s,b_t)>0$ is the proton opacity; physically $\exp(-\Omega(s,b_t))<1$ represents the probability that no inelastic scattering occurs at impact parameter $b_t$. 

The formulae presented above in fact correspond to an over--simplified `one--channel' model, which ignores any internal structure of the proton. This can be readily generalised to a more realistic approach~\cite{Khoze:2013dha} which accounts for the possibility of proton exitation in the intermediate state, $p\to N^*\to p$.  Although we use this latter model for the numerics in this paper, it is in fact the case that, as the photon radiation vertex $p\to \gamma+N^*$ vanishes at $q_\perp\to 0$, while the quasi--on--shell photon transverse momenta $q_\perp$ in exclusive production is very small, the difference between the simplified and more general approach is extremely small for such processes.

\bibliography{references}{}
\bibliographystyle{h-physrev}
\end{document}